\documentclass[aps,amsmath,amssymb,twocolumn]{revtex4}

  \usepackage{graphicx,wrapfig,amsmath,upgreek,mathtools,amsfonts,physics,multirow,color}
  \usepackage{graphicx}
  \usepackage{color}
  \usepackage{amsmath}
  \usepackage{enumitem}
  \usepackage{amssymb}
  \usepackage{hyperref}

\newcommand{\be}{\begin{equation}}
\newcommand{\ee}{\end{equation}}
\newcommand{\ber}{\begin{eqnarray}}
\newcommand{\eer}{\end{eqnarray}}
\newcommand{\cb}{\Box}
\newcommand{\tl}{\triangle}

\begin{document}

\title{Universality in Phyllotaxis: a Mechanical Theory}
\author{Hyun-Woo Lee and Leonid S. Levitov}
\address{
Physics Department, Massachusetts Institute of Technology, 77 Massachusetts Avenue, Cambridge, MA02139}
\date{December 2020}

\maketitle
{\bf Abstract:} One of humanity’s earliest mathematical inquiries might have involved the geometric patterns in plants. The arrangement of leaves on a branch, seeds in a sunflower, and spines on a cactus exhibit repeated spirals, which appear with an intriguing regularity providing a simple demonstration of mathematically complex patterns. Surprisingly, the numbers of these spirals are pairs of Fibonacci numbers consecutive in the series 1, 2, 3, 5, 8, 13, 21, 34, 55... obeying a 
simple rule 1+2=3, 2+3=5, 5+8=13 and so on. This article 
describes how physics helps to clarify 
the origin of this fascinating 
behavior by linking it to the properties of deformable lattices growing and undergoing structural rearrangements under stress. 

\tableofcontents
\medskip 
\noindent {\small
 published in: Symmetry in Plants, Series in Mathematical Biology and Medicine, eds. R. V. Jean, D. Barab\'e, (World Scientific Pub Co Inc, 1998)}

\pagenumbering{arabic}


\section{Introduction}
\label{sec:introduction}
\subsection{The 
patterns of repetitions in plants}
 The visual beauty of plants comes, in part, due to the 
 highly regular and well organized patterns of leaves, florets, seeds, scales and other structural units. 
 Widely admired, the regular patterns in plants are not merely aesthetically pleasing,  they often have 
 unexpected relation with mathematics. 
Here we will be concerned with one particular kind of such patterns --- 
the spirally, or helical, arrangements. 
Spirally patterns occurring in plants have long been known to have a 
surprising 
connection with number theory.  
The area of botany that studies such patterns
is called {\it phyllotaxis} (the word can be translated as
``leaf arrangement''). It has been recognized long ago that
the
so-called Fibonacci sequence: 1, 2, 3, 5, 8, 13, 21, 34, 55, 89
..., where every number in the sequence appears as a sum of two
preceding numbers, is of great importance 
in phyllotaxis. 
The prominence of Fibonacci numbers in phyllotaxis
is well accounted for in both specialized and popular literature, which includes the gems such as
``On Growth and Form'' by
D'Arcy Thompson~\cite{D'ArcyThompson} and ``Symmetry'' by H. Weyl~\cite{Weyl}. 

\begin{figure}[t]
\includegraphics[width=3.4in]{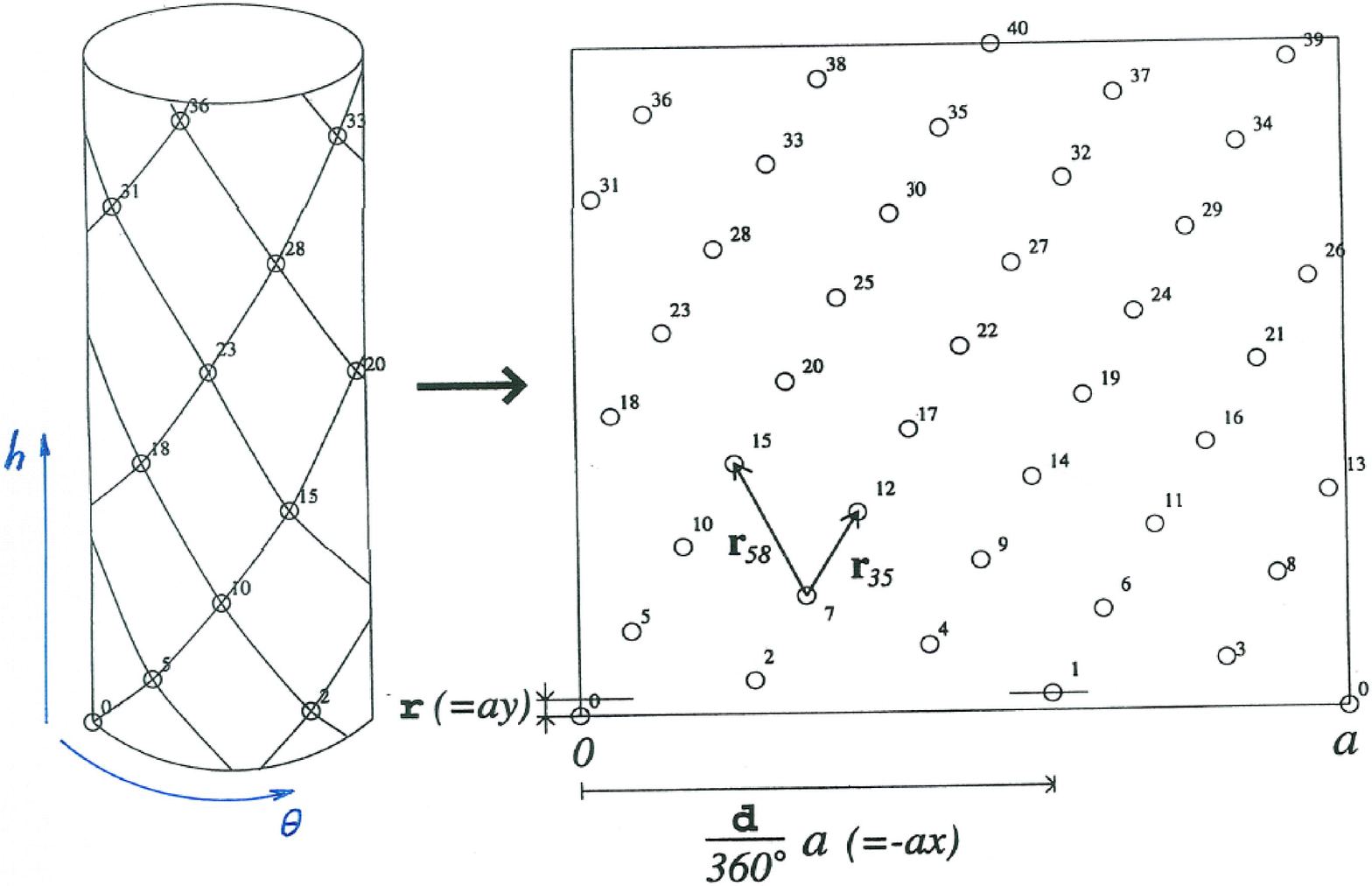}
\includegraphics[width=2.5in]{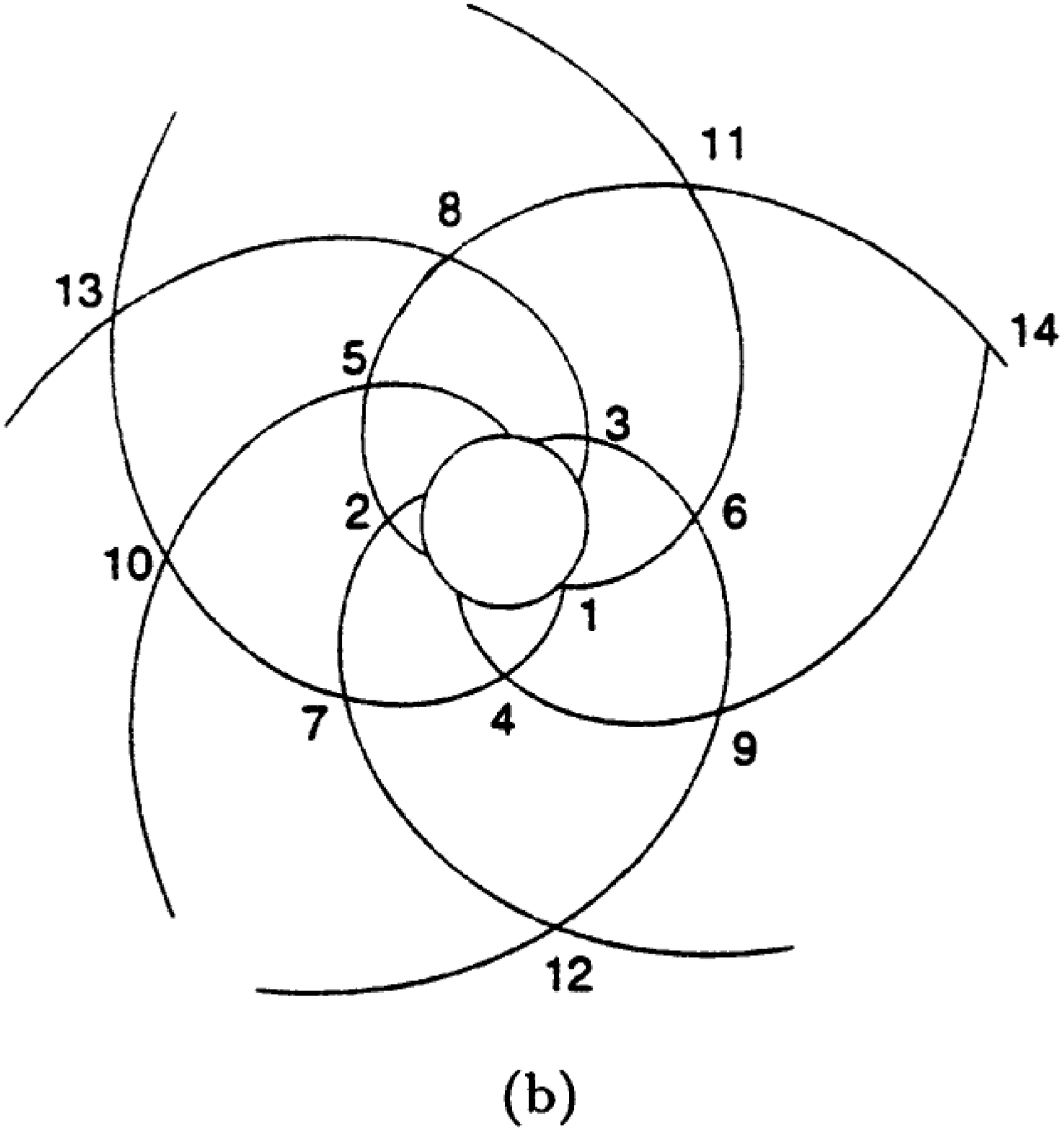}
\vspace{0.15cm}
   \caption{(a) 
Geometric model of a cylindrical plant, such as a pineapple
or a pine cone. The
lines  connecting  neighboring
scales form two families of parastichy helices
with the right and left  helicity.  Each
family contains a Fibonacci number of helices. 
The figure illustrates a model of such a structure by a lattice 
on a cylinder of circumference $a$,
and its mapping onto a periodic lattice in
the plane. The two lattice vectors are shown 
that connect the nearest and
next-nearest neighbors, and define parastichy helices, 
The divergence angle ${\sf d}$ and the helix rise ${\sf r}$
are defined in Eq.(\ref{spiral_lattice}); 
the parameters $x=-{\sf d}/2\pi$ and $y={\sf r}/a$ are defined in
Eq.(\ref{define_xy}). 
 (b) Geometric model of a plant with a disk geometry, such as a sunflower.
The lines connecting neighboring florets generate  two  families  of parastichy 
spirals of the right and left helicity. 
The figure illustrates a model of the floret arrangement by a spiral lattice. The
points on the spirals are numbered according to the distance from disk center.
     }
\label{f1}
\end{figure}

The 
connection between Fibonacci numbers and helical packings of units
in plants, which 
is at the heart of the subject of phyllotaxis, appear to be quite general,
although the details depend somewhat on the plant geometry.
In cylinder-shaped objects, such as fir-tree cones or pineapples, the
scales have a regular arrangement in which 
two families of helices can be identified,
having the right and left helicity. These are 
known in the literature as {\it parastichy helices} \cite{wiki_parastichy} 
(see Fig.~\ref{f1}a). 
The numbers of helices in each family are invariably found to be the 
Fibonacci numbers. 
Moreover, the numbers of the right and left helices 
are non-equal and are given by the pairs of Fibonacci numbers consecutive in the
sequence, such as (2,3), (3,5), (5,8), (8,13) and so on. 
Geometrically, 
such structures can be described as lattices on the surface of a 
cylinder. The appearance of Fibonacci numbers in a
cylindrical structure is called {\it cylindrical phyllotaxis}.

Another type of phyllotactic patterns occurs when units of a
plant are packed on a disk. For example, in a sunflower or a
daisy, the lines connecting neighboring florets define two
families of {\it parastichy 
spirals} \cite{wiki_parastichy}. Again, in each family the number
of spirals is Fibonacci, and in the two families the numbers are
consecutive in the sequence (Fig.~\ref{f1}(b)). For such
structures, it is common to use the name {\it spiral
phyllotaxis}. A detailed discussion of 
the two geometric models of phyllotaxis, 
cylindrical and spiral, can be found 
in the article by Rothen and Koch~\cite{RothenKoch}.

The first studies of phyllotaxis probably go back to the observation made
by Leonardo da Vinci (notebook, 1503) who noted that
  \begin{quote}
Nature has arranged the leaves of the latest
branches of many plants so that the {\it sixth} is always above
the {\it first}, ...
  \end{quote}
(Here $6-1=5$, a Fibonacci number.)
Since then, phyllotaxis has had a long and interesting history,
excellent reviews of which are given, 
e.g., by
Adler~\cite{AdlerI}, Jean~\cite{JeanReview}, and
Lyndon~\cite{Lyndon}. The regularity of the arrangement of leaves
and other similar structures fascinated many biologists,
mathematicians, crystallographers, and physicists, who were
interested in this phenomenon and contributed to the development
of the subject. 
Many renowned scientists contributed to our understanding of phyllotaxis, 
including Kepler, Linnaeus, Bonnet, Bravais,
Airy, 
among others. 
Here, instead of reviewing the historical
development of the subject, we simply summarize, in a roughly
chronological order, the main logical steps of five hunder years
of the research in
phyllotaxis:\\
(a) discovering phyllotactic patterns (XV-XVI century);\\
(b) observing and characterizing (XVI-XVIII century);\\
(c) geometric modeling (since XVIII century);\\
(d) experimental studies (since XIX century);\\
(e) interpreting, explaining (since the late XIX century).

Remarkably, after five centuries of inquiry phyllotaxis
continues to be 
an active research area. The reason 
for that can be seen in 
its multidisciplinary character. It is a problem that does not belong entirely to any particular branch of science, but 
has roots in the subjects as diverse as biology, mathematics, and physics.  
The timeline above makes it evident how these disciplines,
as they progressed, have found application in phyllotaxis by
prompting new questions motivated by their own logic and
providing new valuable perspective on an old problem. 

\subsection{Why Fibonacci numbers?}
\label{sec:origin}
   In the XX century the trends and interests in this field 
   focused
on the question of the origin of phyllotaxis. Here is how the challenge was
stated by a mathematical biologist:
   \begin{quote}
The fascinating question: ``Why does the Fibonacci sequence
arise in the secondary right or left spirals seen on plants?''
seems to be at the heart of problems of plant morphology. In
atomic physics, Balmer's series has opened the way to Bohr's
theory of the atom and then to quantum mechanics and to quantum
electrodynamics. The great hope of bio-mathematicians is that
some day they may be able to do for biology what has been done
by mathematical physicists in physics. (R. V. Jean, Ref.\,\cite{JeanJTB}, p. 641)
\end{quote}
Indeed, despite the long history, 
no general agreement
was reached with regard to the origin of phyllotaxis.
Many explanations proposed in the past
simply state that the Fibonacci packings are optimal in some sense. 
Already Leonardo,
after noticing the regularity in the numbers of leaves, remarks that the
reason they are arranged in such a way probably
has something to do with a better exposure to the sun light through
minimizing mutual shadowing. Later, the logic of the discussion
of the origin of phyllotaxis basically followed that of
Leonardo's, by adding more optimization factors, 
such as air circulation in between the leaves,
or the density of packing of the seeds or florets. The review of
such theories can be found, for example, in the D'Arcy
Thompson's book~\cite{D'ArcyThompson}. 
The  Darwinian evolutionary theory, by asserting that natural selection gives rise to the optimal
structures that are better designed for a local environment, 
apparently 
endows such theories with an air of confidence.

However, such views are known to be at odds with certain experimental observations. 
Indeed, as emphasized in the evolutionary theory, the evolutionary pressure arises from variability within the species. 
Namely, if the species are optimized with respect to a certain factor by evolutionary pressure, there must also occur, 
however rarely, the species that are not optimal but near-optimal.
%
%
If this was the case in the phyllotactic growth, a sunflower that normally exhibits 55 spirals
could sometimes have 54 or 56 spirals. In other words, the
most frequent kind of an exception from the Fibonacci rule
would be associated with the numbers that are close but not equal to the
Fibonacci numbers. 
And yet such numbers
never occur; instead a very different kind of non-Fibonacci patterns is observed. 
The most common exception known to occur is described by the numbers from
the so-called {\it Lucas} sequence~\cite{Weyl}: 1, 3, 4, 7, 11,
18, 29 ... . The sequence is constructed according to the Fibonacci
addition rule, however it starts with a
different pair of numbers. Another aspect of
phyllotaxis that seems to be hard to reconcile with the evolutionary optimization 
theories, is that the non-Fibonacci numbers patterns, even those of the
most common Lucas type, occur extremely rarely, with the probability
of few percent or lower.

The high stability of Fibonacci numbers could make one 
suspect that perhaps they are in some way ``programmed.'' However, 
the conjecture that these numbers are merely encoded genetically seems improbable,
because individuals within the same species often exhibit different
Fibonacci numbers, but hardly ever the non-Fibonacci (Lucas) numbers. Moreover, as in the case of spiral phyllotaxis,
different Fibonacci numbers can occur within one plant.
In addition, by simply making reference to genes one does not come closer to
the understanding exactly what is special about these numbers.
Thus one is led to seek an explanation elsewhere.

In this article we review a mechanical theory~\cite{Levitov1,Levitov2} 
that 
establishes a general physical mechanism of phyllotactic growth. 
Namely, we 
consider the role of mechanical stresses in a 
densely packed arrangement of units in a growing plant.
The stresses, which build up due to the growth anisotropy,
lead to a shear strain in the structure increasing gradually 
and then relaxing by 
a sequence of abrupt structural rearrangements. 
As demonstrated below, this simple
mechanical process gives rise,
exclusively and deterministically, to Fibonacci structures. 

The mechanical explanation of phyllotaxis, 
which links Fibonacci numbers to transitions in the system upon 
stress build-up in the process of growth, is completely general and robust.
In particular, these transitions 
are insensitive to the specific form
of interactions and, occurring one after another, 
generate the entire sequence of Fibonacci structures. Further, besides establishing the 
prevalent character of Fibonacci numbers in phyllotactic patterns the mechanical theory 
explains why the exceptions
are predominantly of a Lucas type. This is achieved by allowing irregularities (e.g. due to fluctuations or noise) to trigger a 
single mistake at an early growth stage.
All predictions of the mechanical theory are therefore in agreement with the
observations.

Besides the mechanical scenario
of phyllotaxis reviewed in this article, 
several other
approaches focused on 
various physical growth-related effects 
that may lead to phyllotaxis. 
Those include theories of growth
mediated by a diffusion of inhibitor 
(Mitchison~\cite{Mitchison}), by a reaction-diffusion process
(Meinhardt~\cite{Meinhardt}), and by mechanical
interactions that control the largest available space (Couder and
Douady~\cite{CouderDouady}). Some of these theories are discussed
in other chapters of this volume. 

Are the mechanisms emphasizing different physical processes mutually exclusive
or complementary to one another?
On a first thought it may 
appear that, given the clear differences between these 
approaches, identifying the right mechanism will invalidate other explanations. 
Yet, below we argue that the situation is considerably more interesting. 
Our analysis of mechanical stresses and their impact on growth 
establishes 
the property of {\it robustness}.  
The robustness 
basically means a wide stability of phyllotactic patterns with
respect to parameter variation in the model. 
The stability property, by an extension, suggests that there is a degree of truth in all theories which invoke 
some form of repulsion/stress during growth --- mechanical, chemical, or else.
In other words, all theories of phyllotaxis which use lattices of objects with some kind of repulsive interactions,
no matter the origin and specific geometry (e.g. cylinder, disk, or cone), 
are equivalent in a ``coarse-grained'' sense. 
The situation here is similar to that in the theory of pattern formation,
where many different microscopic models are known to lead to identical
types of patterns on a larger scale.
In physicist's language, such models, while different in details, belong
to the same universality class.

At the same time, there is an open question of comparing the proposed
scenarios with the underlying physiological processes in biological systems.
This is a fascinating experimental problem that will have to be addressed by future work.
However, whatever the outcome of these studies might be, it is worth noting
that phyllotactic growth occurs in a large variety of biological systems.
It is therefore possible that no unique 
physiological process 
applicable to all systems will emerge, 
and instead several different microscopic scenarios must be considered. 
At the same time, the robustness of phyllotactic growth, as established in the mechanical theory, 
will ensure that the resulting patterns are insensitive to the microscopic details
of the growth.

This article is organized as follows. In
Sec.~\ref{sec:MechanicalTheory} we summarize the mechanical theory of
phyllotaxis and provide a brief review of its history. In
Sec.~\ref{sec:phasespace} we introduce a geometric model that
involves cylindrical lattices and families of parastichy helices, 
defined in terms of the shortest lattice vectors. In this section the notion of 
a phase space of all cylindrical lattices is defined, 
which will be central in our subsequent discussion. Next, in
Sec.~\ref{sec:energymodel} we introduce the energy model, and
consider its symmetries. These symmetries are found to form a large family,
described as an infinite group of modular transformations. Then, in Sec.~\ref{sec:modularsymmetry} the main result of this
article is derived. We use the modular symmetries to relate
different growth stages and thereby
show that Fibonacci numbers are universal in
phyllotaxis. This result, established within the energy model, 
is cast in a rigorous form
of a theorem. From the discussion in
Sec.~\ref{sec:modularsymmetry} 
the robustness property of phyllotactic growth becomes evident. In
Sec.~\ref{sec:robustness} we discuss possible modifications of the model
which can bring it closer to other growth geometries, such as those of spiral phyllotaxis,  
and argue that the stability of Fibonacci patterns and their universality remain unaffected.

\section{Mechanical theory}
\label{sec:MechanicalTheory}
\subsection{Growth 
under stress}
\label{subsec:growthandstress}
The mechanical model discussed here  
involves, in its simplest form, a regular cylindrical lattice of
repulsively interacting objects (see Fig.~\ref{f1}a). Cylindrical lattices provide a convenient representation for a wide variety 
of phyllotactic patterns. Crucially, in this model the lattices are taken to be deformable, with the lattice geometry not fixed rigidly but instead controlled by the balance between external forces and repulsive interactions between different lattice points. 
The repulsion 
can be chosen so that it 
mimics the contact rigidity of the structural units of a plant. However, rather than restricting the repulsive interaction to be a short-range type, it is beneficial to consider a generic repulsive interaction which includes both the short-range and  the long-range parts. 

Further, the effects of growth can be naturally incorporated in the cylindrical lattice model through an external 
force applied along the cylinder axis, which gradually increases as the growth progresses. 
The uniaxial stress due to such a force mimics the stresses arising in the growth of an elongated object in the presence of 
the closed-volume or confinement constraints.
As we will see, under the uniaxial stress the system deforms in such a way that,
starting from a simplest quasi-one-dimensional chain-like structure, it goes through
the sequence of Fibonacci phyllotactic patterns in a completely
deterministic way.

In the mechanical model of phyllotaxis a 
cylindrical lattice deforms 
upon a gradual increase in stress; this deformation is described in terms of a gradually developing shear that tends to minimize the total repulsion energy at each given stress value.
It is crucial, however, that in such a process the structure
does not track the global energy minimum. The  reason  is
that in optimizing its energy the system can explore
only the nearby states by developing small deformations in the lattice.
Therefore, the Fibonacci structures, appearing upon a gradual
increase of the stress,  characterize  the {\it  
progression, or time sequence,} 
of
deformation, 
rather than the result of global energy minimization. 
The Fibonacci structures 
emerge in this framework as the stages of rearrangement
in a deformable system, starting from a certain simple structure and appearing one by one upon 
an increase in the stress during the growth.

In treating the problem 
quantitatively, the first step is to define the mechanical
energy of a lattice using a model interaction
between lattice points, and use this interaction to study
the system evolution under stress (see Secs. \ref{sec:phasespace},\ref{sec:energymodel}). 
This analysis, 
which is straightforward to carry out
numerically~\cite{Levitov1}, 
reveals 
a very robust behavior: for a variety of 
repulsive interactions, 
increasing the stress  drives  the  system  through  all
Fibonacci structures, exclusively and without exceptions.
Further, the model has a distinct advantage in that
it can be treated
analytically  for  a  
large family of interactions (see Sec. \ref{sec:modularsymmetry}). 
In this case  the
robustness and universality  of  Fibonacci  structures  can  be  established 
rigorously~\cite{Levitov2}. 
We subsequently argue (Sec. \ref{sec:robustness}), 
using the robustness property, 
that the results obtained for cylindrical geometry can be generalized to the disk geometry.

Anisotropic mechanical stress accompanying the growth 
is a key ingredient of the 
theory of phyllotaxis 
advocated 
in this article. What can be the origin of such a stress? 
In that regard we note that, while the required stress cannot originate from 
isotropic hydrostatic pressure, 
any growth anisotropy can in general lead
to anisotropic stresses of the form
that generate phyllotaxis.
The details of the relationship between the growth anisotropy and stress
depend on the specifics of the system at hand and, in particular, the system geometry. However, as argued below, these details are inessential for understanding
the general relation between 
the 
stress buildup during growth and the formation of phyllotactic patterns. 




The relation between anisotropic
growth and mechanical stresses is most clear for the 
growth of a cylindrical structure. Consider a cylindrical lattice, such as that pictured in Fig.\,\ref{f1}a, 
which is growing while being 
encapsulated in a fixed volume. Such growth can describe, for instance, the early developmental stages of objects like pine tree cones, which have a
growth center at the apex. At the growth center, new structural units of the
would-be cone are produced at a constant rate, and as a result the cone extends
forward. In free space, such a growth would 
yield one-dimensional
chain-like structures. 
However, if the growth is taking place within a closed volume, the
growing 
cone, or a similar cylindrical object, soon meets a constraint (namely, a boundary) which does
not allow it to freely extend forward. As a result, as more and more units are
generated at the growth center, the cone will deform in order to
fit inside the enclosure. It is instructive to characterize the result
of such deformation by rescaling to fixed density on the cylindrical surface
(see Fig.\,\ref{f1}a). Upon such rescaling, 
the growth can be described as a gradual expansion 
in the direction transverse to the cylinder
axis accompanied by compression along the axis, such that the two-dimensional areal density 
remains fixed.
As the our analysis 
below predicts, such a process leads to, 
exclusively, the 
Fibonacci structures. The actual numbers achieved 
will depend on the growth duration:
the longer the growth continues, the 
higher are the Fibonacci numbers that can be accessed.

What are the implications of the behavior found in cylindrical lattices for other geometries of interest? A useful example that helps to answer this question 
is the growth of a circular object
in a disk geometry, as illustrated in Fig.\,\ref{f1}b.
Such a growth, e.g. describing the development of a sunflower or a daisy, 
has its center at the center of the disk. During the
growth, as more 
units (e.g. florets) 
are added at the center, they push the previously added units, forcing them to 
move outwards along the disk radius. 
One can argue that this process
leads to transformations of the structure equivalent to those of a deformable
cylindrical lattice.

This can be done, e.g., by focusing on the evolution of 
a small rectangular patch of 
the structure, a distance $r$ away from the center. Namely, one can choose the
rectangle sides to be aligned with the
cylindrical coordinates, radial and azimuthal, equal to
$\delta r$ and $r\delta\theta$, respectively, and consider how they change upon growth. As the growth progresses, this 
patch moves radially to a new location $r'>r$, where
the new sides of the rectangle become $(r/r')\delta r$ and $r'\delta\theta$,
since the rectangle
angular dimension $\delta\theta$ 
and its area $r\delta r\delta\theta$ remain constant. 
The rectangle aspect ratio 
$r\delta\theta/\delta r$ increases thereupon by a factor
$(r'/r)^2$. As a result, the part of the lattice within the rectangle,
after moving from $r$ to $r'\gg r$, is
strongly deformed. The deformation degree $(r'/r)^2$ gradually
increases along the radius, whereas the areal density of the florets remains roughly constant. 
Which implies that, since the
main forces in the system are due to nearest neighbors exerting
pressure on each other, locally the mechanics of the deformed spiral
lattice is similar to that of a cylindrical lattice. To map one problem to the
other, 
we can simply replace the spirals of the disk by the
helices of the cylinder, which gives a one-to-one relation
between the parastichy lines connecting the adjacent lattice points. 
One unique aspect of 
the disk growth geometry that distinguishes it from cylindrical geometry is
that 
the deformation varying as a function of radius, as discussed above, gives rise to structural transitions and different phyllotactic domains occurring within the same disk
(see discussion
in Sec.~\ref{sec:robustness}). The concentric annulus-shaped phyllotactic domains have the numbers of spirals (parasticies) which are smaller for the inner domains and greater for the outer domains, increasing with radius (see Sec.~\ref{subsec:spiral}). 
The numbers of spirals, which are Fibonacci numbers, are therefore largest near the disk edge. 

Another geometry of interest 
corresponds to the growth of a plant at an apex 
of its shoot. In this case, the
growth center where new units  of  a  plant  (e.g. leaves, scales, or spines)  
are generated  is
located  at  the  center of the rounded top part of a shoot. 
The  geometry of a curved cone describing such a growth 
can be viewed as intermediate
between the disk  and cylinder geometries discussed above. Near the growth center,
the growth process can be described by the disk  geometry,  whereas some
distance away from it,  as  the  shape  of  the  shoot  is  curved  towards a
cylinder, it resembles the cylindrical growth. 
As in the disk growth case, a thicker shoot gives rise to 
bigger Fibonacci numbers. 

These three basic growth geometries represent different varieties of the general problem of a deformable lattice growing under stress. All three exhibit a very similar behavior. 
While in each case there are specifics 
that lead to some modifications 
in the description of the growth, they do not affect
the stability of Fibonacci numbers. The developmental stability 
follows from the robustness of the sequence of structural changes induced as
the mechanical deformation increases under stress (see Sec.~\ref{sec:robustness}). 
This property is central to understanding the 
universality of Fibonacci patterns. 

\subsection{Early 
mechanical scenarios} 
\label{sec:history}
The first suggestion that mechanical forces 
play a key role in 
phyllotaxis seems to have been made by Hubert 
Airy. In his work ``On Leaf-arrangements'' (Ref.\,\cite{Airy},
p.177)
he wrote:
   \begin{quote}
Take a number of spheres (say oak-galls) to represent
leaves, and attach them in two rows in alternate order (1/2)
along opposite sides of a stretched india-rubber band. Give the
band a slight twist to determine the direction of the twist in
the subsequent contraction, and then relax the tension. The two
rows of spheres will roll up with a strong twist into a tight
complex order, which, if the spheres are attached in a close
contact with the axis, the order becomes condensed into (nearly)
2/5, with great precision and stability. And it appears that
further contraction, with increased distance of the spheres from
the axis, will necessarily produce the order (nearly) 3/8, 5/13,
8/21, etc. in succession, and that these successive orders
represent successive {\it maxima} of stability in the process of
change from the simple to the complex.''
    \end{quote}
This description contains the
key element of the mechanical theory: a spiral  structure  of
rigid  objects that deforms under a stress. However, the way
it is described does not make it clear why  such  a  process
leads  exclusively  to  Fibonacci numbers. 
Moreover, Airy's statement seems to be
more of a conjecture rather than a conclusion, since the analysis
verifying this suggestion has not been done at the time.
To put these ideas on a firm ground 
a theory of the development of a spiral structure 
under mechanical stress is needed.

An important step in 
this direction was made by van Iterson~\cite{Iterson}. 
As a starting point, he considered cylindrical lattices, a geometric model that has been used to describe 
phyllotactic patterns since the work of Bravais~\cite{Bravais}, to which van Iterson added a new ingredient. 
As in the earlier work, a plant 
is represented as 
a cylinder with a spiral arrangement of lattice points on the surface of  a cylinder. In that, the lattice points 
represent individual units of a plant such as leaves, scales, or spines (see Fig.\,\ref{f1}). 
etc. 
The lattice points, if ordered according to their relative heights 
along 
the cylinder axis, $h_m$, $m=0,\pm1,\pm2...$, 
can be described geometrically as a set of 
points on a single {\it generating helix}:
    \be\label{spiral_lattice}
 h_m={\sf r}\,m,\quad \theta_m=\,{\sf d}\,m
,
    \ee
where ${\sf r}$ is the rise of the helix, and ${\sf d}$ is the angular step
known as 
the {\it divergence angle} in the literature. The quantities $h$ and $\theta$, and the lattice points labeled by $m$, are illustrated in Fig.\,\ref{f1}a; here all lattice points are assumed to have different height values $h_m$.
%
The new ingredient of van Iterson's model is identical disks centered at the lattice points, with the arrangement of the  disks 
constrained by the dense packing
requirement. Namely, the disks are packed in a lattice such that each disk is in a direct contact with at least four nearest-neighbor disks which touch but do not overlap. 

\begin{figure}[t]
\includegraphics[width=3.4in]{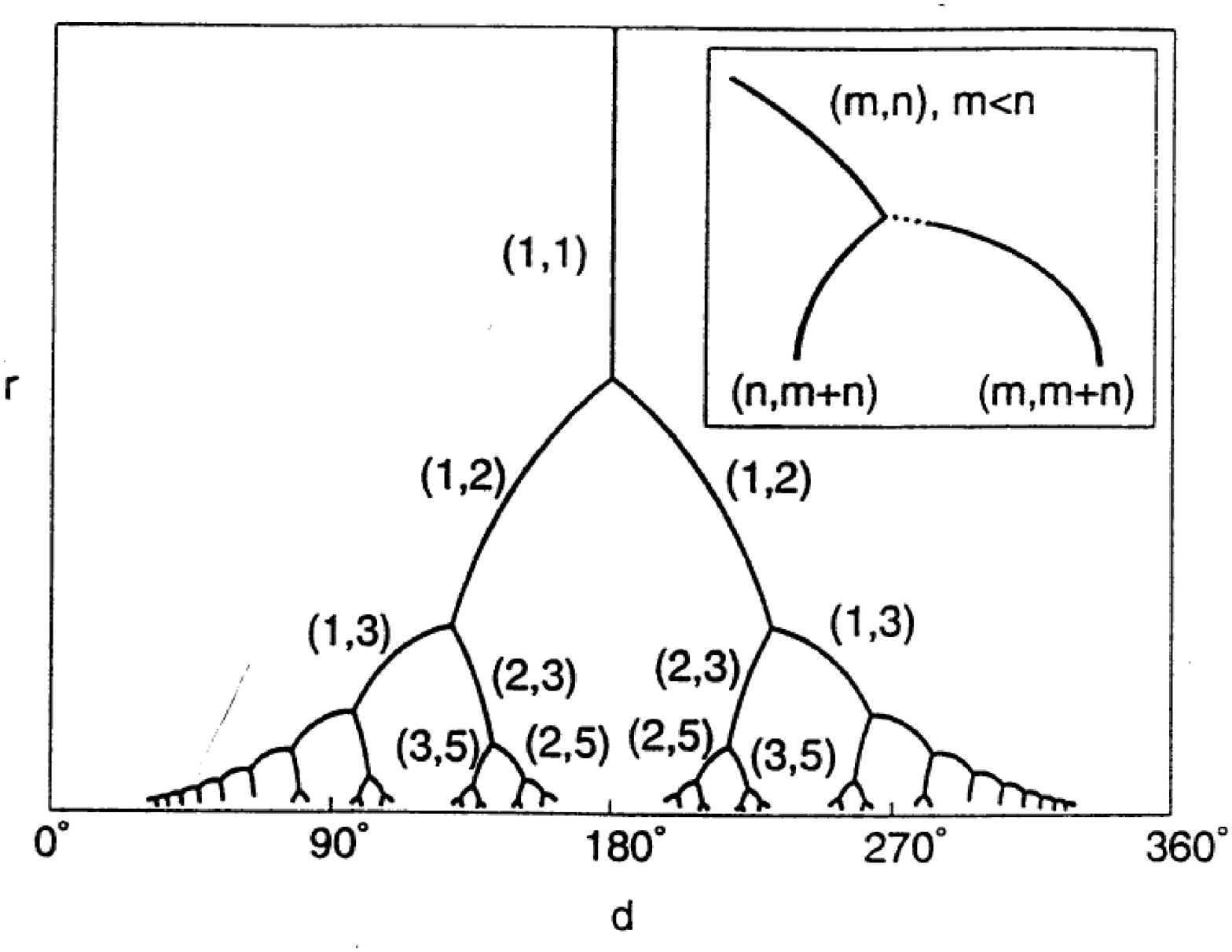}
     \caption{Van Iterson  diagram  of  pairs   $({\sf r},{\sf d})$
corresponding  to  dense  packing  of  disks~\cite{Iterson}.  The diagram has a
structure of a Cayley tree with the branching number $z=3$. 
Parastichy
pairs (the numbers of the right and left helices) are shown next to each arc. \hskip2mm
  Inset:  The contact  pressure theory by Adler~\cite{AdlerII} predicts lifting of the 3-fold symmetry 
at the branching points by opening gaps in some of the arcs. The dotted line represents
unstable disk packings with negative contact pressure.
    }
\label{f2}
\end{figure}

The densely packed disks resemble the densely packed structural units of
a plant, defining the families of right and left parastichy helices 
which connect the nearest-neighbor lattice points as illustrated in Fig.\,\ref{f1}a. The numbers of the helices in the two families 
define the cylindrical lattice ``parastichy type''. (This model, and in particular the relation of the pairs $({\sf r},{\sf d})$ 
to the hyperbolic plane, will be discussed in greater detail in
Sec.~\ref{sec:phasespace}.) 
Van Iterson noticed
that the dense packing requirement constrains the disk arrangements
in a very special way.  
He demonstrated that the lattices of 
densely packed disks, when described in terms of  the generating helix, Eq.(\ref{spiral_lattice}), 
define possible pairs $({\sf r},{\sf d})$ that form a
branching structure known as the Cayley tree (see
Fig.~\ref{f2}). The Cayley tree consists of the arcs of circles
in the $({\sf r},{\sf d})$ plane connected with each other at the 
triple branching points occurring at the ends of the arcs.
The triple branching points
correspond to the maximum-density 
disk packing --- 
the symmetric triangular lattices in which each disk
makes contact with exactly six neighbors. Further, the Cayley tree helps to understand  the properties of the densely packed disk lattices encoded in their parastichy types ---  the numbers of the right and left helices which are uniquely determined by the values of ${\sf r}$ and ${\sf d}$.

These numbers 
show an interesting behavior 
when placed on the corresponding branches of the Cayley tree. 
First, the numbers 
do not change within one branch. Each branch can therefore 
be
labeled by a single pair of integers $(n_1,n_2)$.
This is illustrated in Fig.~\ref{f2} 
where these numbers 
are shown next to the corresponding Cayley tree branches. 
Second, as noted above, the triple branching points correspond to perfect triangular
lattices of maximum density. 
Since in a perfect triangular lattice 
there are three shortest lattice vectors of equal length, and thus three distinct families of helices associated with these vectors, each branching point of the Cayley tree is naturally labeled by three integers $(n_1,n_2,n_3)$. 
As discussed below, these integers obey the identities $n_1+n_2=n_3$, up to a permutation, which might suggests a link to the Fibonacci series. 

Is there a relation between this construction and Fibonacci numbers?
A quick inspection of  Fig.~\ref{f2} indicates that the numbers labeling the arcs are in general non-Fibonacci, with the Fibonacci numbers found only in a small subset of the tree. At the same time, the arcs that are labeled by the pairs of Fibonacci numbers form two continuous paths going from the Cayley tree top all the way down to the bottom. Along these paths the  pairs of
consecutive  Fibonacci numbers are found $(1,2)$, $(2,3)$, $(3,5)...$, 
and  $(2,1)$, $(3,2)$, $(5,3)...$; 
the numbers grow as the rise parameter ${\sf r}$ of the generating helix decreases. 
Taking the decrease in  ${\sf r}$ to be associated with the plant growth  (e.g., see 
Ref.\,\cite{Meicenheimer}), it is tempting to conjecture that 
the arcs of the 
Fibonacci paths 
are somehow linked to different stages of the growth.


These observations were first connected to the effects of mechanical stress
by 
Adler~\cite{AdlerII}.
He considered a force 
that is being applied to a cylindrical
lattice along the cylinder axis and increases gradually. Under growing stress the lattice deforms 
so that the rise of the helix ${\sf r}$  gradually diminishes. 
In the disk model the effect of growing stress
is accounted for by a gradual downward movement along the Cayley tree of the point  $({\sf r},{\sf d})$ that represents the lattice. 
As ${\sf r}$ decreases,
the evolution of the point $({\sf r},{\sf d})$ tracks one of the tree branches
until a branching point is reached. At which point the system must choose between the
two arcs going further down. If we assume that the system, represented by the point
$({\sf r},{\sf d})$, would always choose, by some mechanism, to evolve
from the branching point 
along the arc marked by a pair of Fibonacci numbers, the 
model would predict phyllotactic growth. However, in the original formulation of
the disk model it was unclear what mechanism could guide the growth in
the right direction at the branching points.

Adler proposed to replace the purely geometric condition of a
dense packing of the disks by a more physical notion of an
external pressure which is applied to the system of rigid disks and is balanced
by the contact pressure between the disks (see also
~\cite{Schwendener}). The contact pressure theory ``resolves'' the
triple points on the Cayley tree diagram; this happens because the condition for mechanical equilibrium predicts that 
one of the arcs that meet at each branching point
represents a system with negative
contact pressure. The negative pressure regions correspond to
unstable structures and must be eliminated from the diagram. This leads 
to gaps opening in the arcs which 
divide the tree into
separate paths of arcs that run downward (see inset of Fig.~\ref{f2}). The evolution along each of these paths, owing to the absence of branching, is
deterministic. Further, the two Fibonacci paths discussed above 
remain unaffected by the negative pressure gaps. 
Now, 
as ${\sf r}$ diminishes, the system can evolve
through the sequence of Fibonacci states in a perfectly deterministic fashion. 
(Further details of this construction are discussed 
in the chapter by Adler in this volume.)

To summarize this discussion, by  introducing  contact  pressure  between  the  disks   Adler
completed and solved van Iterson's disk model. His solution provided
a proof of phyllotaxis for the contact-pressure interaction. While realistic, 
this interaction  
is not found in any biological
system. Moreover, since the phenomenon of phyllotaxis is
so widespread, and the form of interaction presumably varies strongly among different species, 
a solution for a particular
interaction only partially addresses the problem,
leaving the question of robustness 
unanswered. Proving the 
universality of phyllotaxis 
requires 
establishing the
stability of Fibonacci growth in a sufficiently generic class of
models.

This is precisely what is accomplished by the energy model~\cite{Levitov1,Levitov2}. The energy model is essentially
a generalized contact pressure model, with the contact 
interactions between disks replaced by a generic repulsive interaction between points of a cylindrical lattice.
A big advantage of the energy model is that it cleanly delineates the physics
ingredients (interactions) from the geometry ingredients (cylindrical
lattices). This makes it possible 
to treat the problem of phillotaxis in full rigor and generality.  These results will be
summarized and reviewed below.
Quite unexpectedly, the separation of interactions and geometry reveals
that the problem possesses a hidden symmetry, namely the modular
symmetry group $GL(2,Z)$. With the help of the  $GL(2,Z)$ symmetry the occurrence of Fibonacci numbers in phyllotactic patterns can be established for generic repulsive interactions. This analysis demonstrates the universality of Fibonacci phyllotaxis, providing an explanation of the Fibonacci growth.

Further insight into the mechanical origin of phyllotaxis was provided by an experimental work published 
recently by Couder and Douady~\cite{CouderDouady} (see their
chapter  in  this  volume).  They  devised a 
hydrodynamic  system  that  models spiral phyllotaxis with the help of magnetically  polarized droplets of a  ferrofluid on an oil surface. In this
experiment~\cite{CouderDouady}, the ferrofluid droplets
appear  in  a  regular  sequence  at  the middle of an oil disk,
representing units of a phyllotactic pattern  generated  at  the
growth  center.  The  droplets repel each other by the dipole $r^{-3}$
interaction; 
this gives rise to an  effective  pressure
from the newer droplets on the older ones that organizes them in
Fibonacci  patterns  similar  to  those  seen in the spiral
phyllotaxis. This work proves  that,  as  envisioned  by  Airy,
mechanical   forces   are indeed sufficient  to  produce  phyllotactic
patterns.

In addition to the non-biological realization in the oil drop experiment, 
it was proposed that phyllotaxis can be observed in other
physical systems, such as cells of Benard
convection~\cite{RivierEtAl} and flux lattices in
super\-con\-ductors~\cite{Levitov1}.


\section{The geometric model}
\label{sec:phasespace}
\subsection{Cylindrical lattices and parastichy helices} 
\label{subsec:cylindricallattices}

In this section we review in greater detail the cylindrical lattice model of phyllotaxis. 
This model provides, in particular, a 
definition of the lattice parastichy type given by the numbers of the right and left helices. 
In our discussion 
we will focus on the 
dependence of these quantities on lattice geometry.
Further, we will
establish a relation between the 
phase space of cylindrical lattices and the hyperbolic plane. This
relation will prove quite useful later in the analysis of the
energy model. Our discussion of cylindrical lattices overlaps in part
with Refs.~\cite{Erickson,Rivier,RothenKoch}.

Cylindrical lattices can be described by mapping them to lattices
in 
a plane. 
For a given cylindrical lattice
(\ref{spiral_lattice}) the planar lattice is obtained by unrolling the cylinder as a wallpaper roll. Upon unrolling 
a two-dimensional lattice is obtained 
that has periodicity given by the cylinder circumference. 
Namely, working in Cartesian coordinates 
%
\be
\label{eq:cartesian_system}
{\bf r}=u {\bf i}+v{\bf j}
,
\ee
we align the cylinder axis parallel to the ${\bf j}$ axis and roll it along 
the ${\bf i}$ axis direction. 
Accordingly, each point of the generating helix (\ref{spiral_lattice}) is mapped on
a one-dimensional array of points parallel to the ${\bf i}$ axis, 
${\bf r}_p={\bf r}_0+pa{\bf i}$, with $a$ 
the
cylinder circumference. In this way we obtain 
a two-dimensional periodic lattice:
   \ber
{\bf r}_{pm}  =  u_{pm}{\bf i}+v_{pm}{\bf j} =
a\left((p-mx){\bf i}+my{\bf j}\right)
                         \nonumber \\
 =  \sqrt{A}\ \left(\frac{p-mx}{\sqrt{y}}{\bf i}+ m\sqrt{y}{\bf j}\right),
\label{lattice}
    \eer
where $p$ and $m$ are integers, $A=a{\sf r}$ is the unit cell area
(see Fig.~\ref{f1}a).
The first line of Eq.(\ref{lattice}) gives the planar lattice obtained by unrolling the cylinder; in the second line 
new parameters are introduced that will be used throughout our discussion below. These are 
the relative row-to-row displacement $x$ of the generating helix, 
Eq.(\ref{spiral_lattice}), and the
height-to-circumference ratio $y$,
related to the parameters ${\sf r},\,{\sf d}$ as
   \be\label{define_xy}
x=\,-{\sf d} /(2\pi),\quad y=\,{\sf r}/a
.
   \ee
The helical lattice
(\ref{spiral_lattice}) is completely specified by three
parameters ${\sf d}$, ${\sf r}$, and $a$. Accordingly, the lattice
(\ref{lattice}) is characterized by the parameter values $x$, $y$, and $A$.

The representation involving the 
quantities $x$ and $y$, which is used
in our discussion below, 
has a number of advantages, some obvious and some less obvious. 
First, this representation is scale-independent (i.e. is invariant upon rescaling),
since $x$ and $y$ depend only on the angles between the basis vectors 
of the lattice and their relative sizes, 
but neither on the cylinder circumference $a$ nor the lattice unit cell area $A$.

Further, as demonstrated below in Secs.~\ref{sec:energymodel},\ref{sec:modularsymmetry}, 
these parameters present an intrinsic advantage 
from a geometric viewpoint, helping to link cylindrical lattices to hyperbolic geometry. 
Namely, the lattice geometry 
can be described by a single {\it complex parameter} 
\be
z=x+iy
\ee
that can be treated as a variable in the hyperbolic plane.
The energy of the lattice will be shown to be invariant under
modular transformations 
\be
z\to z'=(az+b)/(cz+d)
.
\ee 
This result will be crucial for our analysis
of hidden symmetries of the lattice energy in Sec.~\ref{sec:modularsymmetry}, 
and for establishing 
stability of the Fibonacci phyllotaxis. 

In a lattice (\ref{lattice}) 
the parastichy helices 
are introduced with the help of shortest lattice vectors. In a generic
lattice, each point has two nearest
neighbors, where ``nearest'' refers to the metric in the Cartesian $(u,v)$
plane (\ref{eq:cartesian_system}), 
$\Delta=\sqrt{(u-u')^2+(v-v')^2}$. 
Alternatively, the distance can be 
measured on the curved cylinder surface using geodesics. This defines, up
to a sign, a lattice vector ${\bf r}_{pm}$ connecting each lattice 
point to its nearest
neighbor. Similarly, there is a pair of next-nearest neighbors
that defines a vector ${\bf r}_{qn}$. Generally, $|{\bf
r}_{qn}|>|{\bf r}_{pm}|$, however, for the lattices with a
rhombic unit cell the lengths of two vectors are equal. The
vectors ${\bf r}_{pm}$ and ${\bf r}_{qn}$ will be called {\it the
pair of shortest vectors}.

Given ${\bf r}_{pm}$, for any lattice point ${\bf r}$ we can define the corresponding 
{\it parastichy line} 
by drawing a straight line through ${\bf r}$
and the nearest lattice points 
${\bf r}\pm {\bf r}_{pm}$. This gives a line ${\bf r}+s {\bf r}_{pm}$ with $s$ a real-valued parameter, which contains an array of lattice points 
$\{{\bf r}+k{\bf
r}_{pm}\}$, where $k$ is an integer. On the cylinder, this line
corresponds to a helix drawn through ${\bf r}$ and connecting it to 
nearest neighbors. 
Different helices 
obtained in this way 
for different
${\bf r}$'s form a {\it parastichy family}. 
Likewise, the second parastichy family is defined by 
connecting lattice points in the next-shortest vector
${\bf r}_{qn}$ direction. The {\it parastichy
numbers} $P_1$, $P_2$ which define the lattice parastichy type are 
defined as the numbers of the helices in these two
families (see also Refs.~\cite{Erickson,Rivier,RothenKoch}). 

The definitions above are of course all but natural as the helices on a cylindrical plant picked by 
a human eye do connect nearest neighbors.  
In the remaining part of this section we review 
several useful properties of the shortest lattice vectors and their relation with the numbers $P_1$, $P_2$. Namely, for the shortest and next-shortest vectors ${\bf r}_{pm}$ and ${\bf r}_{qn}$ given by (\ref{lattice}) the numbers $P_1$ and $P_2$  equal, up to permutation, $|m|$ and $|n|$.

First, we show that the  
vectors ${\bf r}_{pm}$ and ${\bf r}_{qn}$ are primitive vectors, i.e. they provide a basis for
the lattice (\ref{lattice}). Although this property is nearly obvious, we sketch a quick proof both for completeness and as a reminder of the fundamental property of the lattice unit cell area that will be useful below. 

To prove that the shortest vectors ${\bf r}_{pm}$ and ${\bf r}_{qn}$ form a
basis, it is sufficient to show that they define a parallelogram
of a minimum area corresponding, in terms of the areal density, to exactly one lattice point, $A={\sf r}a$.  Suppose the latter were not true, then the
parallelogram spanned by ${\bf r}_{pm}$ and ${\bf r}_{qn}$, besides the lattice points at the vertices, would
have contained an inner lattice point. 
We arrive at a contradiction by noting that the shortest distance from 
any inner point of a parallelogram to its vertices is smaller than one of the parallelogram sides $|{\bf r}_{pm}|$ or $|{\bf r}_{qn}|$.

As a side remark, the shortest vectors provide what may be viewed as a natural basis.  Because of the shortest vector property the parallelogram spanned by ${\bf r}_{pm}$ or ${\bf r}_{qn}$ has a minimum size compared to those for other pairs of primitive vectors. 
As a quick reminder, 
the choice of a basis in a lattice is not unique,
and any pair of primitive vectors 
can serve as a basis. 
For example, the lattice (\ref{lattice}) 
has a basis defined by primitive vectors
\[
{\bf e}_1=a{\bf i},\quad {\bf e}_2=-{\sf r}x{\bf i}+{\sf r}y{\bf j}
.
\] 
While this is a legitimate basis for this lattice, 
the lengths of the vectors ${\bf e}_{1,2}$ can greatly
exceed the distance between the nearest lattice points in the $(u,v)$ metric.
To the contrary, the basis formed by the shortest vectors ${\bf r}_{pm}$ and ${\bf r}_{qn}$ 
would be optimal (and thus natural) in the sense of this metric, in loose analogy with the role 
of the Wigner-Seitz cell in studying crystal lattices.



Next, we establish the relation between the vectors ${\bf r}_{pm}$ and ${\bf r}_{qn}$ coordinates and the parastichy numbers, given by $P_1=|m|$ and $P_2=|n|$ (up to a permutation). 
We first analyze the family of parastichy lines obtained with ${\bf
r}_{pm}$. 
Consider the domain 
in the $(u,v)$ plane 
(\ref{eq:cartesian_system}) defined by
$0\le u<a$ and $0\le v<v_{pm}={\sf r}m$.
This is a 
rectangle of area $mA$ which contains exactly $|m|$ lattice
points including ${\bf 0}$. Because these points are not nearest
neighbors of each other in the sense of ${\bf r}_{pm}$, any parastichy line passing through one of them
does not pass through any other of these points. On the other hand, two
opposite sides of the parallelogram are related by a translation
by ${\bf r}_{pm}$, which means that each parastichy must pass
through one of these points. Therefore, the number of 
parastichies generated by ${\bf r}_{pm}$ is $|m|$. By the same argument, the
number of parastichies in the other family, generated by ${\bf r}_{qn}$, 
equals $|n|$.

Since the shortest vectors ${\bf r}_{pm}$ and ${\bf r}_{qn}$ are
defined up to a sign, we can assume, without loss of generality, 
that both $m$ and $n$ are positive
integers. With this convention, used thoughout the article, 
the parastichy numbers are just $P_1=m$ and $P_2=n$.

\subsection{The space of cylindrical lattices}
\label{spaceoflattices}
    Here we consider the relation between the parastichy numbers $P_1$,  $P_2$ and the parameters $A$, $x$, $y$ of cylindrical lattices.
This relation, as will shortly become clear, 
is central to understanding 
the geometry of phyllotactic
patterns. Our analysis 
will demonstrate a nontrivial
relation between the quantities  $P_{1,2}$ and the hyperbolic plane parameterized by
the complex variable $z=x+iy$. 
The numbers $P_{1,2}$ will be shown to be encoded in a 
``hyperbolic wallpaper'', consisting of domains in the hyperbolic plane that are 
mapped on each other by 
modular transformations.
This connection will provide 
a tool that will help us 
to explain Fibonacci phyllotaxis and understand its stability.

The cylindrical lattices
(\ref{lattice}) are completely defined, up to a rescaling factor $\sqrt{A}$, 
by specifying the parameters $-\infty<x<\infty$ and $0<y<\infty$. 
Hence the upper halfplane $y>0$ of the $(x,y)$ plane 
can serve as the phase space of all such lattices.
The rescaling factor will be irrelevant for most of our discussion.
Thus, unless stated otherwise, we focus on the 
lattices of unit density, $A=1$.

The pairs $(x,y)$ are in one-to-one correspondence with 
the planar lattices (\ref{lattice}), 
since any such pair defines a lattice and vice versa.
For the lattices on a cylinder, however, 
all pairs 
$(x+n,y)$ with integer $n$ 
correspond to the same lattice, because changing $x$ by an
integer amounts to changing ${\sf d}$ in (\ref{spiral_lattice})
by a multiple of $2\pi$, which obviously does not change the
generating spiral. Hence $x$ can be chosen in the interval $0\le
x<1$. Furthermore, the transformation $x\to 1-x$ maps
a cylindrical lattice with one helicity to an equivalent lattice
with the opposite helicity, right-hand to left-hand or vice versa.
Therefore, all non-equivalent lattices can be parameterized by $0.5\le x<1$ 
(or, equivalently, by $0\le x<0.5$) 
up to interchanging the right-hand and the left-hand
helicity. Below
we adopt this convention in all drawings and figures. Later, however,
when 
discussing the energy of a lattice in
Secs.~\ref{sec:energymodel},~\ref{sec:modularsymmetry},~\ref{sec:robustness},
it will be more convenient to consider the entire halfplane $y>0$
in the $(x,y)$ plane, because
the transformations $x\to n\pm x$ are merely a 
subgroup
of a 
symmetry group of the energy function.

We now proceed to discuss how the $(x,y)$ plane is organized by the parastichy pairs
$P_1$, $P_2$.
Given a particular parastichy pair, $P_1=m$ and $P_2=n$, what is the set
of $x$ and $y$ for which it is realized? We will call this set a
{\it parastichy domain} in the $(x,y)$ plane corresponding to the parastichy
pair $(m,n)$. (The scale-invariant definition of the parastichies 
makes the rescaling factor $\sqrt{A}$ an irrelevant parameter.)
To find the shape of the parastichy domains, let us assume
that the pair of shortest vectors ${\bf r}_{pm}$ and
${\bf r}_{qn}$ is fixed, and study how this restricts $x$ and $y$.

One requirement for ${\bf r}_{pm}$ and
${\bf r}_{qn}$ comes from the fact that the
parallelogram spanned by these vectors must have area $A$,
since it is a unit cell of the lattice (see above).
Therefore we must have $|{\bf r}_{pm} \times {\bf r}_{qn}|=A$. 
This condition is equivalent to
   \be\label{determinant}
pn-qm=\,\pm 1\ ,
   \ee
which restricts possible combinations of $p$, $m$, $q$, and $n$.
First, it follows from (\ref{determinant}) that the integers $m$ and $n$ are
mutually prime. Also, if $m$ and $n$ are positive and not equal 
to $1$ simultaneously,
the integers $p$ and $q$ are either
both positive, or both negative.

Given the $m$ and $n$ values, what can be said about $p$ and $q$?
Different pairs $(p,q)$ corresponding to the
same
$(m,n)$, satisfying condition (\ref{determinant}), are related by the transformation $p'= p+km$, $q'= q+kn$, where $k$
is an integer. According to Eq.(\ref{lattice}), the change
$(p,q)\to(p',q')$ can also be realized by a transformation
of the lattice parameter $x'=x+k$. Combining it with
the transformation $x'=1-x$, changing helicity, we see that the
pair $(m,n)$ uniquely determines $p$ and $q$
for $x$ within the interval $0.5\le x<1$.

\begin{figure}[t]
\includegraphics[width=3.4in]{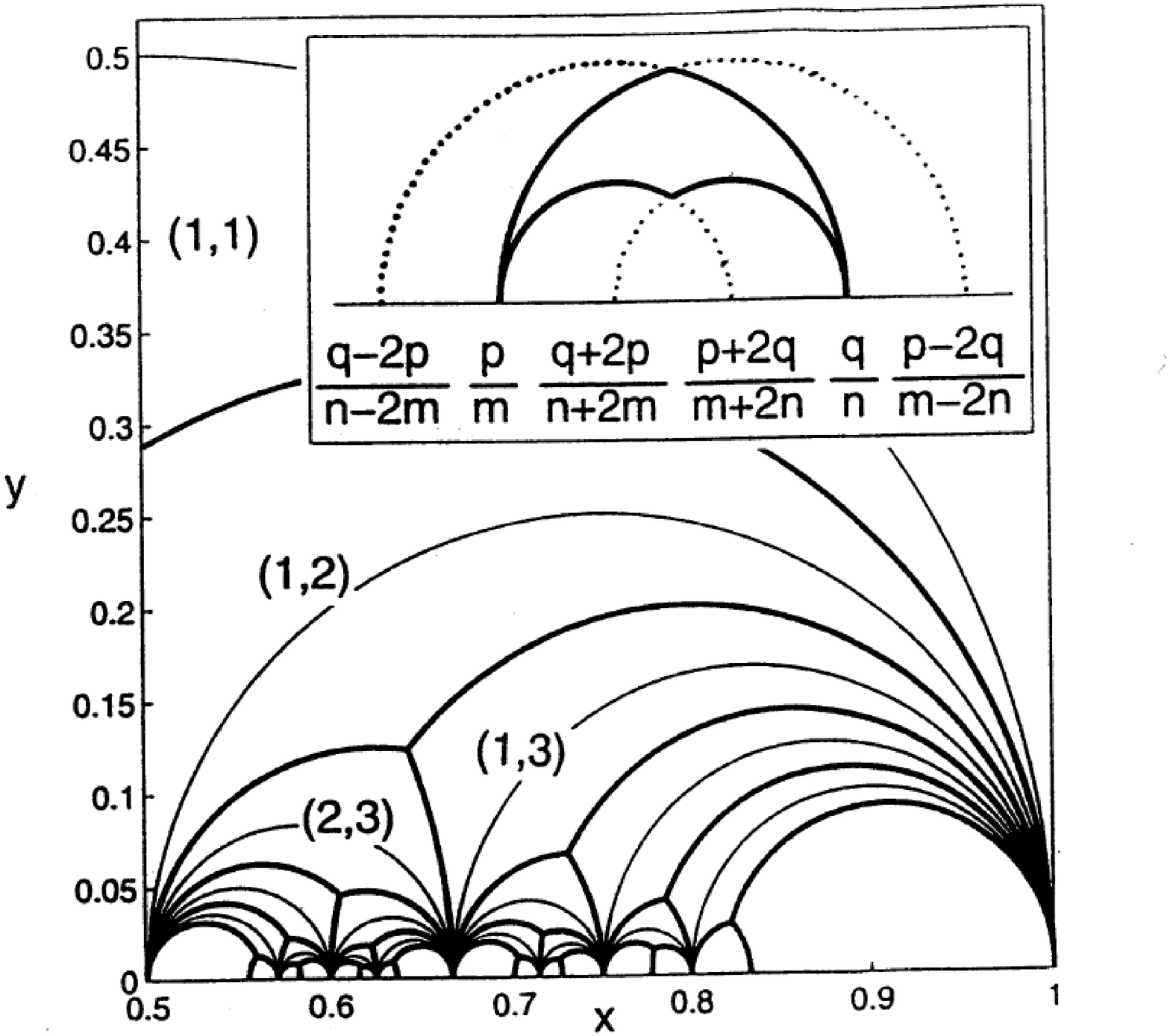}
    \caption{The $x,\,y$
plane partitioned by parastichy domains (thick lines).
The domains are marked by parastichy pairs $(P_1,P_2)$.
Only the  $0.5<x<1$ region is shown, since the pattern
is symmetric under $x\rightarrow x\pm 1, x\rightarrow 1-x$. 
Farey triangles are shown by thin lines (see Fig.\,\ref{f6}).
\hskip3mm
  {\it Inset:} A parastichy  domain  bounded  by  four
semicircles (\ref{eqn:parastichydomain}).}
\label{f4}
\end{figure}

Now let us consider how $x$ and $y$ are constrained by the condition 
that ${\bf r}_{pm}$ and ${\bf r}_{qn}$ are the shortest
vectors. Since each of the diagonals of the parallelogram spanned by 
${\bf r}_{pm}$ and ${\bf r}_{qn}$ must in this case be longer than
its sides, we can write four inequalities:
\be
{\rm (a)}\ |{\bf r}_{pm} \pm {\bf r}_{qn}| > |{\bf r}_{pm}|;
\quad
{\rm (b)}\ |{\bf r}_{pm} \pm {\bf r}_{qn}| > |{\bf r}_{qn}|.
\ee
To see what these conditions mean for $x$ and $y$ let us write them explicitly
using the form (\ref{lattice}) of the lattice
vectors. The condition (a), taken with both the plus and minus signs, 
restricts the pair $(x,y)$ to be in the region
\ber\nonumber
&& (n+2m)\left((x-x_1)(x-x_2)+y^2\right) > 0,\quad
\\ \label{eqn:parastichydomain}
&& (n-2m)\left((x-x_3)(x-x_2)+y^2\right) > 0,\quad
\\ \nonumber
&& {\rm where}\  x_1 = {q+2p \over n+2m}
        ,\  x_2={q \over n},\ x_3={q-2p \over n-2m}
\eer
%
The condition (b) takes a similar form. Taken together, (a) and (b) define a
curvilinear domain in the $(x,y)$ plane, as shown in the inset of
Fig.~\ref{f4}. The domain boundaries are arcs of semicircles
with the diameters on the $x$ axis. For any $x$
and $y$ inside the domain the lattice (\ref{lattice}) is described
by the parastichy pair $P_1=m$, $P_2=n$.

Similar domains can be obtained for all combinations of $p$, $m$,
$q$, and $n$, which satisfy the condition (\ref{determinant}).
These domains partition the $(x,y)$ plane as shown in
Fig.~\ref{f4}. In this figure we display the parastichy domains
only in the region $0.5<x<1$ because
our partition of the plane
is invariant under translations $x\to x+ k$
and reflections $x\to k-x$.
This invariance property accounts for the fact that
the pairs $(x,y)$, $(x+1,y)$, and $(-x,y)$ represent the same lattice up to helicity change.

There is a useful relation of the plane partition by the
parastichy domains $P_1=m$, $P_2=n$ and van Iterson
diagram (Fig.\,\ref{f2}). The boundaries of the parastichy domains
correspond to rhombic lattices, because at the borderline where
the parastichy pair changes the lattice must have next-shortest vectors of
equal length. Therefore, the points where the domains join in
three correspond to perfect triangular lattices (this property wil be
discussed further in Sec.~\ref{sec:fareypartition}). At the same time,
these points are the triple branching points of van Iterson
tree. Thus each branch of the tree resides within a single parastichy
domain, connecting its opposite corners 
(compare Fig.\,\ref{f2} and Fig.\,\ref{f4}).
After associating each van Iterson branch with a parastichy domain
we see that the parastichy numbers in van Iterson diagram are in fact 
identical to those introduced above using the plane partition.
Therefore, each branch of van Iterson diagram is characterized by
parastichy numbers $P_1$, $P_2$ which are constant within it
and can change only at the branching of the tree. 

It is 
interesting to understand the organization of
the parastichy numbers $P_1$, $P_2$ throughout the $x,\,y$ plane in Fig.\,\ref{f4}. Inspecting the parastichy pairs written in the adjacent domains shown Fig.~\ref{f4} inset gives a simple addition rule: 
For a
domain with $P_1=m$ and $P_2=n$, the two domains adjacent to it
from below (closer to the $x$ axis) are lebeled by the pairs 
$P_1=m+n,\,P_2=n$ and
$P_1=m,\,P_2= m+n$. As illustrated in Fig.\,\ref{f4}, this relation between parastichy pairs of
neighboring domains, 
so far conjectured empirically, describes organization of pairs $P_1$, $P_2$ in the entire $x,\,y$ plane. 
The addition rule for the numbers $P_1$, $P_2$ will be proven in Sec.~\ref{sec:modular_symmetry}. 

For now taking the addition property
for granted, 
we can infer that for {\it any} pair of mutually
prime integers $m>0$ and $n>0$ a domain 
labeled by such a
pair can be found in Fig.~\ref{f4}. 
It thus becomes evident that the pairs of consecutive
Fibonacci numbers correspond to a very small sub-family of
parastichy domains, while most of the domains are non-Fibonacci.

This observation helps to 
clarify that the cylindrical lattices 
realized in the natural world 
are a very small subset of all lattices
that 
are mathematically possible, adding suspense to the problem of explaining the widespread occurrence of Fibonacci numbers. 
To help demystify it, in the next section 
additional ingredients will be added in the discussion: mechanical forces,
energy and the development under stress.


\section{The energy model}
\label{sec:energymodel}
\subsection{Phyllotactic growth under stress: the 
sequence of deformations and adjustments}
    To describe interaction between different units of a
phyllotactic pattern, we employ a repulsive potential $U({\bf
r})$ of a generic form~\cite{Levitov2}. 
In our model, this potential defines forces by which the points constituing 
the cylindrical lattice (\ref{spiral_lattice}) repel each other.
From a microscopic point of view, the interaction $U$ models the effects
of rigidity of structural units and of contact pressure between them, as well as other similar effects arising at short distances 
(e.g. the volume constraint in dense packing of scales of a pine-tree cone). 
While the interactions of a short range or hard core type perhaps would be the most relevant
for systems of interest, it is beneficial to allow for a generic repulsive interaction, since this helps to assess the robustness of phyllotactic growth.

Next, we introduce the interaction potential describing forces between the lattice points and define the energy functional. To that end, we will focus on the central force potential model,
in which the force ${\bf F}=-\nabla U(r)$ 
points along the line connecting two interacting points. 
The potential $U(r)$ in this case 
is a function of the distance measured
using the Euclidean metric in the plane obtained by unrolling 
the cylinder and the cylindrical lattice into 2D, as discussed above. To define the energy functional for a lattice we first note that the total energy obtained from all pairwise interactions, 
given by the expression
    \be
 E_{\rm total}={1 \over 2} \sum_{pmp'm'} U(|{\bf r}_{pm}-{\bf r}_{p'm'}|),
\label{eqn:totallatticeenergy}
     \ee
is formally divergent for an infinite lattice, either or a cylinder or in 2D. 
It is therefore more natural to consider the `energy density' defined as energy per one lattice site. This quantity is given by\footnotemark
    \be
 E(x,y)= \sum_{pm} U(r_{pm}),
\quad
r_{pm}=|{\bf r}_{pm}|,
\label{eqn:latticeenergy}
    \ee
where the sum runs over all
vectors of the lattice (\ref{lattice}). The quantity (\ref{eqn:latticeenergy}) is equal to the energy density per unit area times the unit cell area $A$. Hereafter we suppress the dependence
on $A$, focusing 
mainly on the constant density case, $A=1$. We note parenthetically that 
the ${\bf r}=0$ term can be taken out from the sum (\ref{eqn:latticeenergy}) with no impact on the discussion. Indeed, this term corresponds to self-interaction, and thus eliminating it would merely 
change the energy by a constant which is independent of the lattice geometry and thus inessential for our analysis.

Below we will focus on the case of repulsive interactions, $dU(r)/dr<0$. In addition, for mathematical convenience, we assume that that $U(r)$ decays
rapidly enough to assure convergence of the sum (\ref{eqn:latticeenergy}).
The functional form of the interaction $U(r)$ can be taken, for example, as an expeonential $e^{-r/r_0}$, a gaussian $e^{-r^2/r_0^2}$, or a power law $r^{-\gamma}$.
It will be clear from our discussion that the qualitative behavior is independent of the particular form of interaction.

One caveat associated with our definition of energy, Eq.(\ref{eqn:latticeenergy}), is that this quantity is defined 
as a sum over {\it all} lattice points (\ref{lattice}), of which the 
points ${\bf r}=u{\bf i}+v{\bf j}$ with equal $v$ coordinates correspond to the same point of the spiral structure
on the cylinder (\ref{spiral_lattice}). Despite this, we treat these points as distinct in the sum over $i$ and $j$. Such a choice of the 
expression for the energy, Eq.(\ref{eqn:latticeenergy}), is deliberate: as we demonstrate below, the quantity
(\ref{eqn:latticeenergy}) has hidden symmetries which 
underpin the robustness and universality of phyllotactic growth. 

Further, one can argue that the approximation made in replacing the energy of a cylindrical lattice by that of a lattice in the two-dimensional plane is better than it sounds. Indeed, comparing the 
energy (\ref{eqn:latticeenergy}) to the energy of the cylindrical
structure (\ref{spiral_lattice}), with the interaction defined 
using the shortest distance on a cylinder, we note that the difference of the
two energies corresponds to the part of the sum (\ref{eqn:latticeenergy})
with the $u$ component of the radius vector exceeding half of the cylinder
circumference, $|u|>a/2$. However, as we will see below, 
the development under uniaxial stress makes the circumference $a$ increase 
so that already at a relatively early stage of growth
it can exceed the range of interaction set by $U(r)$. As soon as this happened, 
the inaccuracy in
the expression (\ref{eqn:latticeenergy}) becomes insignificant.
This observation justifies using the expression (\ref{eqn:latticeenergy})
instead of a marginally more accurate but less symmetric expression for cylindrical lattices (\ref{spiral_lattice}).

Next, we proceed to analyze the transformations of cylindrical lattice structures 
under stress. Increasing the pressure acting 
along the cylinder axis has two distinct effects. One is a compression of the system in the $z$-axis direction. Another is a buildup of pressure accompanied by a density change in the system. For simplicity, below we will treat these two effects as decoupled, assuming that the
variation of stress does not affect pressure in the system, and
hence the lattice density $\rho=A^{-1}$ is constant. This approximation is made on the grounds that the system compressibility depends mostly on the lattice average density and not as much on the details such as the angles between nearest-neighbor bonds and other geometric details~\cite{LeeLevitov}. We will therefore treat the distinction between the conditions of constant pressure and constant density as inessential, ignoring it in our discussion.

We implement the constant density approximation by describing the lattice defomration through changing the
lattice parameter $y$ and, at the same time, maintaining constant $A$. As $y$
varies gradually from higher to smaller values, for each given $y$ value we
have to adjust $x$ for the lattice energy to attain a local
minimum,
   \be
{\partial E\over\partial x}=0,\quad
{\partial^2 E\over\partial x^2}>0,\quad A,\, y=\text{const}
\label{minimuminx}
    \ee
and  then  to  examine  the  evolution  of  optimal  $x$  as $y$
decreases from infinity to zero. The process  (\ref{minimuminx})
makes  $x$  an  implicit  function  of $y$. One can say that the
function $x(y)$ describes the progression (or, ``history'')  of  the  deformation.
That  is, we assume that $y$ is controlled externally, and $x$ is
a free parameter in which the system is trying  to  reach  local
equilibrium.  The  reason the roles of $x$ and $y$ are different
becomes obvious from their geometric meaning: $y$ controls  the
spacing  of  lattice  points  along the cylinder axis, while $x$
corresponds to a lattice shear, or to a twist of the cylindrical
structure (since $A$ is constant, neither $x$ nor $y$ affect the lattice density).

\begin{figure}[t]
\includegraphics[width=3.5in]{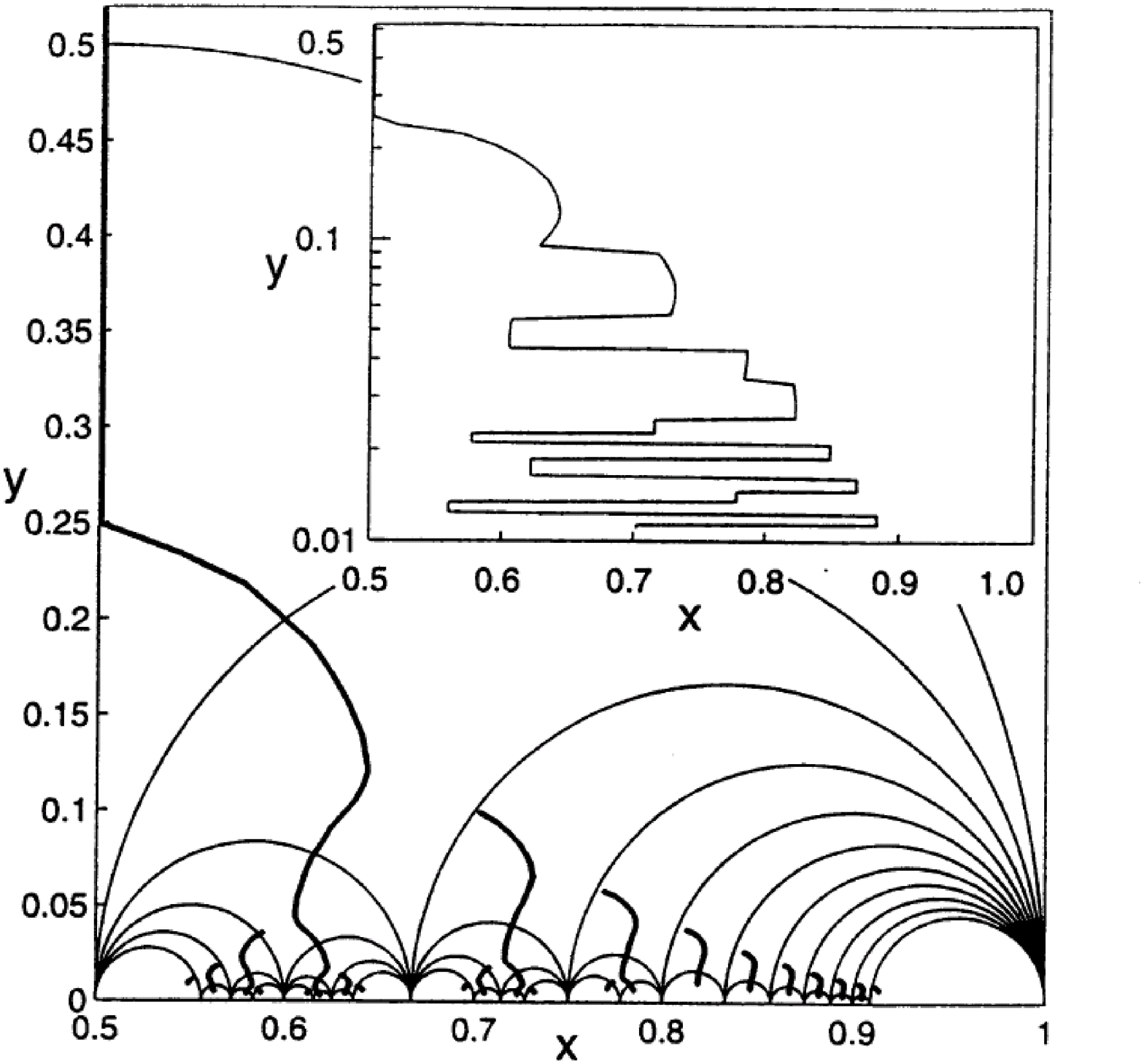}
\caption{
The dependence $x$ {\it vs.} $y$ (thick  lines) 
for the local minima of the energy (\ref{eqn:totallatticeenergy}), 
describing cylindrical lattice (\ref{lattice})
evolution under increasing uniaxial stress.
The interaction $U(r)=e^{-5r}$ and unit density, $A=1$, was used 
in calculation.
Farey triangles  (Fig.\,\ref{f6}) are shown 
to help exhibit the structure of the pattern (thin lines).
\hskip3mm
  {\it Inset:}
The trajectory of the global minimum of the energy (\ref{eqn:totallatticeenergy})
switches erratically between different
branches of the local minima shown in the main panel.
   }
\label{f5}
\end{figure}

The energy functional $E(x,y)$ governs the deformation and shear developing in the compressed lattice. The resulting trajectory in the lattice phase space $(x,y)$ can be linked to the behavior of the energy minima, Eq.(\ref{minimuminx}). 
Namely, as the parameter $y$ is becoming smaller under uniaxial compression, the energy
$E(x,y)$, taken at a fixed $y$ value, acquires more and more extremal points in $x$.
In the phyllotaxis problem we are interested in tracing out the (local)
minimum towards which the structure evolves without jumping to
other minima. As we will see, it is instructive to analyse all local
minima on equal footing. Such analysis will provide, in particular, an insight into the structure of the patterns resulting from the minima evolution under variation of $y$. This is exemplified in Fig.~\ref{f5} where the trajectories for different energy minima obtained for the
interaction $U(r)=e^{-5r}$ are shown. The resulting pattern of trajectories has a number of interesting properties:
\begin{enumerate}
\item   
For $y$ large enough there is  only  one  energy  minimum, located 
at $x=0.5$, which corresponds  to  the  angle  $\alpha$  in  the
generating spiral (\ref{spiral_lattice}) equal  to  $180^{\text{o}}$.  In terms of  cylindrical lattices, $x=0.5$ describes the  simplest imaginable structure in which all lattice points reside in one plane that cuts the cylinder vertically in two halves, the even-numbered and odd-numbered points located on the opposite sides of the cylinder in an alternating order.\\
\item   
At $y\approx 0.25$ this structure becomes unstable, and acquires
a twist with a right-hand or left-hand helicity. The two opposite-helicity states
are related by mirror symmetry, and thus
have equal energies. At this $y$ value a single energy minimum transforms into a pair of adjacent minima. This is manifest in the $x=0.5$ trajectory splitting up at $y\approx 0.25$, and can be associated with a bifurcation of the evolving structure.\\
\item   
As $y$ is lowered further, trajectories in Fig.~\ref{f5} display no other branching or bifurcation
below the first bifurcation encountered at $y\approx 0.25$. This is despite the fact that many new energy minima appear as $y$ decreases, however each of these minima appears some
distance away from the existing minima.\\
\item   
There are two {\it principal} trajectories that emerge at
the bifurcation. Because of what has been said, the principal
trajectories display smooth evolution below the bifurcation, displaying no secondary bifurcations. Hence, provided that the
system evolution goes through the bifurcation point, at subsequent
compression it will follow one of the two principal
trajectories.
\end{enumerate}
    Of course, while the detailed behavior of the trajectories depends somewhat on the choice of the interaction $U(r)$, the general behavior of
the minima appears to be robust. In particular, the properties 1,2,3,4
highlighted above are insensitive to the particular form of the interaction. This is  exemplified, for instance, by the analysis in Refs.~\cite{Levitov1} and
~\cite{Levitov2}, which focusses on the interactions 
$U(r)=\exp(-r^2)$ and $U(r)=-\ln r$, respectively.

Next we proceed to discuss what  numbers of spirals, i.e. the parasticy numbers, can be realized for the structures obtained for different trajectories. This can be analyzed most easily by superimposing the trajectories with the parastichy domains constructed above in 
Sec.\ref{spaceoflattices}
After superimposing the trajectories with the domains in Fig.\ref{f4}, 
one makes a striking observation that {\it the principal trajectory, following the  bifurcation,
passes  exclusively through  the Fibonacci domains.} This result
means that, starting from the  simplest  structure, the system does not have 
any other choice but to evolve into a Fibonacci structure. 
The system, responding to $y$ gradually varying by maintaining
local equilibrium in $x$, is driven through the Fibonacci sequence of states. This is so  because
in  order  to  switch  to a non-Fibonacci state it has to overcome a
finite energy barrier.

We note that in our approach to modeling phyllotaxis 
by a requirement that the system traces local energy minima 
the word ``local'' is absolutely crucial. If instead one would have chosen to analyze the {\it global} energy minimum, for each
$y$ seeking the $x$ value yielding the lowest energy state, the result
would have been quite different. This is illustrated in the inset of
Fig.~\ref{f5} which shows the position of the global
minimum which jumps in a perfectly erratic way
between the principal and non-principal trajectories. Such jumping trajectory shows no regularity whatsoever, in particular it passes through many non-Fibonacci parastichy domains and generates structures of a mostly non-Fibonacci type. 
This obviously means that the global energy
criterion, while being useful in a variety of other problems, does not provide a good guidance in the phyllotaxis problem. One might argue that this is in a sense natural since the phyllotactic systems of interest, which develop into Fibonacci structures, are macroscopic even at an early stage of growth. Indeed, 
even an embryo at an early developmental stage is a 
macroscopic object which is unlikely to transcend structural barriers due to thermal or environmental fluctuations, even if that would take it to a lower energy state. 
This is obviously related to the fact that for a macroscopic system the time required for reaching the lowest
energy state by jumping over barriers would be very large, presumably much larger than the growth
time. We therefore conclude that, while local energy minimization has a clear physical significance in describing growth, the models based on a global energy minimization, which are blind to the presence of barriers, are of limited utility. 

Returning to the discussion of diffrent trajectories in Fig.~\ref{f5}, it is intresting to note 
that  the  {\it   next
principal}   trajectory   in  Fig.~\ref{f5} 
generates structures from the
Lucas sequence 1, 3, 4, 7, 11, 18, etc. This is in good agreement with (and provides an explanation for) the well known fact that the Lucas numbers are the most
common  exception  in phyllotaxis. To obtain these numbers through our mechanical development model one simply has to assume that
the system makes one mistake at the very beginning by jumping over  the
energy  barrier  to  the  lattice  in  the  $(3,1)$  domain, 
after which it strictly follows the rules of the game.  One  can crudely
estimate  the barrier height, and see that it increases
inversely with $y$. This implies that a mistake, if
happened at all, would be most likely to occur at higher $y$ values, i.e.
at the beginning of the growth. In other words, the Lucas sequence is associated with the mistake in the growth which is the most likely one to occur. 

We verified, by performing numerical simulations and otherwise, that the results described above show considerable robustness and are not interaction-specific. Namely, we find that all `reasonable' repulsive interactions, $dU/dr<0$, fit the bill. 
One  might argue  that a  particular  form  of  the
repulsive  interaction would  not matter as long as it renders the lattice stable. This conjecture is indeed true, as will be discussed in the next section where we show that  the  energy
model  can  be  treated  analytically and rigorously. After describing the rigorous results,  we will  return to the
robustness property and formulate more precisely the conditions
on the potential under which the energy model leads to Fibonacci
structures. The robustness property provides a lot of freedom in
varying the form of interaction. Furthermore,  one  can  generalize  the
results  to  the  interactions  that vary during the growth (see
Sec.~\ref{subsec:cylindricalphyllotaxis}),   which provides insight into 
phyllotactic growth in the geometries  other than cyclindrical (see
Sec.~\ref{subsec:spiral}).

Finally, we make a cautionary remark that using the degree of compression $y$
as  a  control  parameter  may  not be entirely physical. In a real
system, the external forces producing stresses in a growing system correspond to pressure and strain, with the resulting  deformation governed  by the conditions  of
mechanical  stability. However, it can be shown~\cite{LeeLevitov}
that in our problem the stress and the  deformation  are  in  a
one-to-one relation, and so, to avoid unnecessary complications,
in what follows we will replace the actual external forces by the
quantity $y$ defined above, which will act as a control parameter .

\subsection{Symmetries of the energy}
\label{sec:modular_symmetry}
In this section we will discuss the symmetry properties of the energy
function, Eq.(\ref{eqn:latticeenergy}), focusing on the topography defined by $E(x,y,A)$ in the $x-y$ plane. As we will see, the behavior of $E(x,y,A)$ is rather peculiar: there are infinitely many energy
minima which are all degenerate, i.e. correspond to identical energy values. Furthermore, the $E(x,y,A)$ topography is such that the minima are organized in an intricate network resembling a mountain range with a system of valleys surrounded by peaks and passes. To understand the resulting structure, we will introduce a
group of modular symmetries, comprised of the transformations of the $x-y$ plane that leave the function $E(x,y,A)$ invariant. We will choose the fundamental domain of the symmetry group and define the corresponding partition of the $x-y$ plane such that it elucidates the network induced by the $E(x,y,A)$ topography. As we will see, this construction greatly facilitates the analysis and leads to a simple explanation of the pattern of the
growth trajectories, such as that in Fig.~\ref{f5}.
In subsequent 
sections the modular symmetry will be used to treat the
problem analytically and rigorously, and to prove the stability of Fibonacci
structures in a rather general way.

To make our discussion of modular symmetry more transparent, it will be convenient to introduce the
complex variable $z=x+iy$ and view the 
$(x,y)$ plane as a complex $z$ plane. The energy function $E(x,y,A)$ will then be defined on the so-called modular space, allowing our analysis to benefit from a high symmetry revealed by such a construction. Accordingly, we will use the notation $E(z,A)$ instead of  $E(x,y,A)$. 

Furthermore, 
it is also convenient to use complex parameterization for the `physical space' (the unfolded cylinder) where the lattices (\ref{lattice}), obtained by unrolling 
cylindrical lattices, are defined. In passing to the complex notation we identify ${\bf i}\to1$, ${\bf j}\to i$, after which Eq.(\ref{lattice}) reads
   \be\label{complexlattice}
r_{pm}(z)=u_{pm}(z)+iv_{pm}(z)
=\sqrt{A\over{\rm Im}(z)}\ (p-m\overline{z})\ ,
   \ee
where $\overline{z}=x-iy$.

General {\it modular transformations} of a complex plane are defined in a standard way as 
fractional linear transformations of a complex variable. 
These trasformations and the associated mappings of the complex plane are often encountered in the complex-variable calculus and its applications~\cite{Apostol}. 
In particular, these transformations play an important role in 
the hyperbolic geometry. This relation, as will become clear shortly, will be pivotal for our discussion. 

Integer modular transformations of a complex
variable~\cite{Apostol} are defined by using a unimodular
$2\times2$ matrix with integer elements,
    \be
{\bf A}=\left( \begin{array}{cc}
                  a & b \\ c & d
            \end{array} \right), \quad |\mbox{ det } {\bf A}|=1\ .
    \ee
Then the corresponding modular transformations of $z$ are defined as
    \ber\label{eqn:modularsymmetry}
     z \rightarrow z' & = & (az+b)/(cz+d) , \quad
         \mbox{if\ \ det }{\bf A}=1 , \nonumber \\
     z \rightarrow z' & = &
        (a\overline{z}+b)/(c\overline{z}+d), \quad
         \mbox{if\ \ det }{\bf A}=-1 \ .
    \eer
Separate definitions for the two signs of the determinant 
$\text{det} {\bf A}$ are required to assure that the halfplane $\mbox{ Im }z'>0$ remains invariant under these transformations. 
The transformations in Eq.(\ref{eqn:modularsymmetry}) define an analytic
function for $\text{det}\,{\bf A}=1$, and an anti-analytic function for
$\text{det}\,{\bf A}=-1$. 

It is easily verified that the transformations in Eq.(\ref{eqn:modularsymmetry})
form a group, with the group multiplication 
represented by matrix multiplication. Namely, the composition of two modular transformations $z\to z'\to z''$ is a modular transformation associated with a matrix ${\bf A' A}$, where ${\bf A}$ and ${\bf A}$ describe the transformations $z\to z'$ and $z'\to z''$, respectively.

In group theory, the $2\times 2$ integer matrices with a unit determinant and a group multiplication defined through matrix multiplication is known as the group $SL(2,Z)$. Here we are interested in a bigger group known as $GL(2,Z)$ comprised of integer matrices with the determinant $+1$ or $-1$ which contains the group  $SL(2,Z)$ as a subgroup, $GL(2,Z)/SL(2,Z)=Z_2$. 

The significance of these transformations 
is elucidated by the following

\noindent
{\bf Theorem:} The energy $E(z,A)$ is invariant under any
modular transformation (\ref{eqn:modularsymmetry}).

\noindent
{\bf Proof:} As a first step, we show that under the transformations of $z$
given in Eq.(\ref{eqn:modularsymmetry}) the lattices (\ref{complexlattice}) change in avery simple way. Namely, these transformations define Euclidean rotations of the lattice, when $\mbox{ det }{\bf A}$=1, and a rotation combined with a mirror reflection, when $\mbox{ det }
{\bf A}=-1$. 


Indeed, by a direct calculation one verifies that the lattice
(\ref{complexlattice}) changes under the transformations
(\ref{eqn:modularsymmetry}) as follows:
     \ber
    r_{pm}(z) & = & e^{i\phi}r_{p'm'}(z')\quad
        \mbox{for\quad det }{\bf A}=1, \nonumber \\
    r_{pm}(z) & = & e^{i\phi}\overline{r_{p'm'}(z')}\quad
        \mbox{for\quad det }{\bf A}=-1,
\label{eqn:rotation}
     \eer
where $p'$, $m'$ and the rotation angle $\phi$ are defined by 
     \ber
p'=ap+bm, \quad m'=cp+dm, \label{eqn:helicsnochange} \nonumber \\
\exp(2i\phi)=(c\overline{z}+d)/(cz+d).
     \eer
Eq.(\ref{eqn:rotation}) means that the lattices
(\ref{complexlattice}), under the transformations
(\ref{eqn:modularsymmetry}), are mapped to isometric lattices.

Next, to complete the proof, we note that the lattice energy $E(z,A)$, Eq.(\ref{eqn:latticeenergy}), is given in terms of an isotropic, angle-independent interaction $U(r)$. 
As a result, the quantity $E(z,A)$ is invariant under distance-preserving transformations of the lattice, such as the rotations and mirror-reflections in Eq.(\ref{eqn:rotation}).\\ \hfill {\bf QED}

Symmetry can be used to gain insight into the properties of
energy $E(z,A)$ in a much the same way as, for
example, periodic functions are analyzed by studying them within the fundamental
period of the translation symmetry group and then using the periodicity property to extend the function to
the entire space. To apply this strategy we need to identify a suitable tesselation of the complex $z$ plane induced by the $GL(2,Z)$ symmetry. This tesselation will play a role in our analysis analogous to periodic tesselations of space used for describing the structure of periodic functions.

An appropriate group-theoretic vehicle used to characterize functions invariant under some
symmetry group is that of a {\it fundamental domain}.
The fundamental domain is defined a geometric shape in the space of the action
of the group such that its images obtained by applying all
group elements cover the whole space with no gaps and no overlaps 
(except, possibly, over boundaries). Loosely speaking, the
fundamental domain is a minimal space region that, after having been
replicated by the group transformations, partitions the entire space.


For the symmetry group (\ref{eqn:modularsymmetry}), we
need a domain in the $z$ plane chosen so that the mappings
(\ref{eqn:modularsymmetry})
of
the domain cover the $z$ plane, overlapping only along
boundaries.
The fundamental domain construction for the modular group, 
introduced by Gauss, is widely discussed in 
mathematical literature.
(For example, see Ref.\,\cite{Apostol}, Chap. 2, and Ref.\,\cite{hyperbolicgeometry},
Chap. 5.)
However, the standard
Gaussian fundamental domain known as the ``modular figure'' will not be that
useful, because for our purpose it is too small. Instead, we will use a
3 times larger domain, which is not a truly fundamental domain in the
sense of the standard definition. We will see below that the larger domain is
more
natural from the point of view of the energy topography in the $z$ plane.
The relation of our domain to the Gaussian domain will be
discussed in Sec.~\ref{sec:fareypartition}.

Let us begin with introducing in the halfplane ${\rm Im}\,z\ge0$
a family of semicircles:
  \be\label{triangleside}
y=\sqrt{\left(x-{p\over m}\right)\left({q\over n}-x\right)}\ ,\
{p\over m}\le x\le{q\over n}\ ,\
  \ee
where $p$, $m$, $q$, and $n$ are integers with $pn-mq=\pm 1$ and
$m>0$, $n>0$. (In Figs.~\ref{f4},~\ref{f5} the semicircles are shown by thin lines.)
We will denote each semicircle by its end points:
$[p/m, q/n]$. It is convenient to allow formally the
combinations in which either $m$ or $n$ is zero, by adding
vertical straight lines $x=q$. (It follows from $pn-mq=\pm 1$
that if $m=0$, then $n=1$.) Our notation for such ``generalized
semicircles'' that connect $x=q$ with $x=\infty$, will be $[1/0,
q/1]$.

It turns out that two semicircles of this family can intersect
only at the real axis, at the point where they are tangent. By
virtue of this property, the semicircles divide the $x-y$ plane
into curvilinear triangles with the vertices on the $x$ axis at
the points $x=p/m,\ q/n,\ r/s$, where $r=p+q$, $s=m+n$.
We will denote such
triangles by specifying the $x$-coordinate of their vertices:
$[p/m, q/n, (p+q)/(m+n)]$. These triangles are called {\it Farey
triangles}, and are closely related to the so called Farey
numbers. Farey triangles and Farey numbers constitute a nice subject on the
borderline between arithmetic and elementary geometry,
and are well accounted in mathematical literature
(see, e.g., the textbooks~\cite{Apostol,hyperbolicgeometry}.)

\begin{figure}[t]
\includegraphics[width=3in]{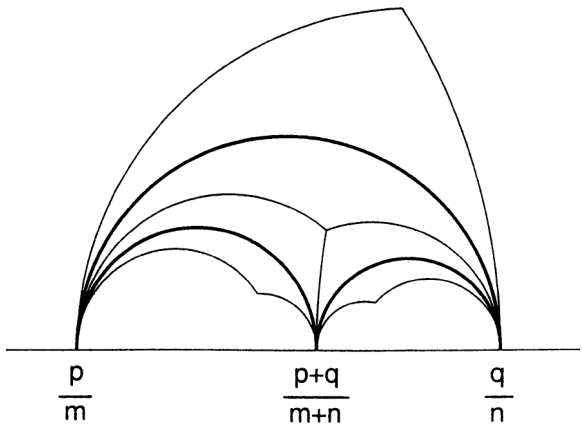}
\caption{
  Relation between the Farey triangles (Fig.~\protect\ref{f4}, thick lines)  
and  parastichy  domains  (Figs.~\protect\ref{f4},\protect\ref{f5}, thin
lines), which partition the complex  $z$ plane in two different ways,
is illustrated.
Each Farey triangle is centered at a corner of three adjacent parastichy
domains. Each parastichy domain ovelaps with two adjacent Farey triangles.
   }
\label{f6}
\end{figure}

In the number theory, Farey numbers is a name for a
construction that organizes all rational numbers $0\le p/m\le 1$
in a hierarchy. Starting with $0=0/1$ and $1=1/1$, one applies
the Farey sum rule, $p/m\oplus q/n =(p+q)/(m+n)$, and
successively generates more and more rational
numbers (see~\cite{Apostol}, Chap.5). The order in which the numbers are
generated coincides with the hierarchy of our Farey triangles:
one vertex of each triangle is Farey sum of two other vertices
(see Fig.~\ref{f6}).

Let us comment on the relation between the Farey triangles and
the parastichy domains. Both partitions of the $z$ plane are
invariant under modular transformations. Comparing
Eq.~\ref{triangleside} with the parastichy domains construction
of Sec.~\ref{spaceoflattices}, it is obvious that for a given
Farey triangle $[p/m,q/n,(p+q)/(m+n)]$, the sides $[p/m,q/n]$,
$[p/m,(p+q)/(m+n)]$, and $[q/n,(p+q)/(m+n)]$ belong to the
parastichy domains with the parastichy pairs $(m,n)$, $(m,m+n)$,
and $(n,m+n)$, respectively (see Eq.~\ref{eqn:parastichydomain}
and Fig.~\ref{f6}). Hence, each Farey triangle
overlaps with three parastichy domains, one side per one domain,
and conversely, each parastichy domain overlaps with two Farey
triangles. The relation of the two ways of partitioning the
$z$ plane will be studied in more detail in Sec.~\ref{sec:modularsymmetry},
and then used for the discussion of the stability of Fibonacci
numbers.

\begin{figure}[t]
\includegraphics[width=3.5in]{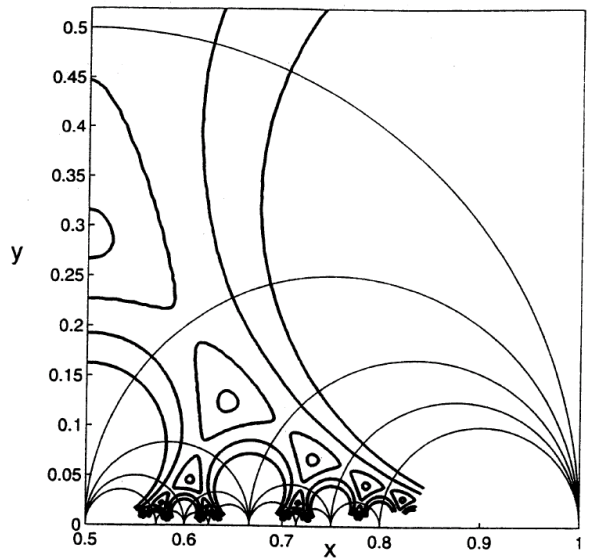}
\caption{
   A contour plot (thick lines) of the lattice  energy  for  the
interaction  $U(r)=\exp(-5r)$.  Farey triangles (thin lines) are
shown for guidance. The contours behavior in all Farey triangles
is similar.
   }
\label{f7}
\end{figure}

To see the role of Farey triangles, in Fig.~\ref{f7} we draw
a contour plot of the energy for the potential
$U(r)=e^{-5r}$, together with the Farey triangles
partitioning the $z$ plane. It is evident from the figure
that the energy contours display similar behavior in each
triangle. Qualitatively, from the topography point of view,
there is a ``valley'' inside each Farey triangle, with a minimum
at the center. For each valley there are three neighboring
valleys in the adjacent triangles, ``connected'' with it by
``passes'' through the saddles located at the middle of each
side of the Farey triangles. At the corners of the triangle
there are peaks of infinite height.

Schematically, the valleys in Fig.~\ref{f7} are connected with
the neighboring valleys in a structure that can be characterized
as a Cayley tree with a branching number three. This tree can be
compared to the pattern of trajectories in Fig.~\ref{f5} obtained
by looking at the energy minima in $x$ at fixed $y$. Obviously,
no matter what $y$, the minima will reside somewhere within the
valleys, away from the peaks. Therefore, when $y$ varies
continuously, the minima trace out the valleys and their
interconnections. Comparing Fig.~\ref{f5} to Fig.~\ref{f7}, it is
evident that the pattern of the minima approximately repeats the
pattern of the valleys. From a topological point of view, the
branching points of the Cayley tree of valleys correspond to
quasibranching of the minima trajectories that one defines by
bridging different trajectories across the gaps~\cite{Levitov2}.
As $y$ varies, each trajectory explores a sequence of
neighboring valleys, and quasibranchings manifest that there are
three neighboring valleys around each valley.

However, let us emphasize that the trajectories of minima in
Fig.~\ref{f5} are {\it not invariant} under the modular
transformations. Technically speaking, the reason is that we are
looking for minima of a modular symmetric function $E(z,A)$
under a non-symmetric constraint: $y={\rm const}$. Obviously, if
the pattern of minima were symmetric, all parastichy pairs would
be equally likely to appear. Hence, the absence of symmetry is
an ultimate cause of the appearance of Fibonacci numbers.

It is important to realize that the problem has incomplete
symmetry: the lattice energy is modular-symmetric, but the
deformation process is not. Thus a concept of {\it asymmetry}
emerges, which implies the absence of symmetry, but nevertheless
certain closeness to an exact symmetry. We will see that the
asymmetry is a much more powerful notion than just total absence
of symmetry. The modular symmetry transformations enable one to
compare the energy minima trajectories in different Farey
triangles. It is even correct to say that the problem of
stability of Fibonacci numbers in phyllotaxis is now reduced to
the analysis of the asymmetry of the pattern of energy minima
trajectories. This idea will be the bottom line of the following
discussion.


\section{Universality of Fibonacci numbers}
\label{sec:modularsymmetry}

\subsection{Plan of the discussion}
\label{sec:discussion}

In this section we will prove a rather general  statement  about
any  energy  minima trajectory. Although, strictly speaking, we
are interested  only  in  the  principal  trajectory  (which  is
Fibonacci),  there  is  a  lot  of  advantage in considering all
trajectories together.

To formulate our main result, let us recall that a
generalized Fibonacci sequence is a sequence of integers
$\Phi_n$ obeying the Fibonacci recursion relation,
   \be
\Phi_{n+1}=\Phi_n+\Phi_{n-1},
   \ee
where  the  starting  numbers $\Phi_0$, $\Phi_1$
can   be   any integers.  For example, the standard Fibonacci sequence
and the Lucas sequence are given  by  $\Phi_0=1,\,\Phi_1=1$  and
$\Phi_0=1,\,\Phi_1=3$, respectively.

{\bf Main theorem:} For any interaction $U(r)$ from the family
defined below, a continuous trajectory of the energy minima in
the $z$ plane goes through the sequence of parastichy domains with
the parastichy pairs from a generalized Fibonacci sequence. The
first two numbers of the sequence are determined by the
beginning of the trajectory.

Once the theorem is proven, the result about having only
Fibonacci numbers on the principal trajectory  follows from
the fact that it begins in the $(1,1)$ parastichy domain.

It is worth remarking that in the theorem the word
``continuous'' is crucial! Given that the deformation of the
lattice is continuous, the theorem states that the deformation
stages all are (generalized) Fibonacci. On the other hand, if
the deformation makes the lattice unstable, and it abruptly
transforms to some other structure or lattice, with a jump on
the $z$ plane, the theorem is not claiming anything.

The class of potentials $U(r)$ for which the theorem holds
is defined by the following two conditions.\\
  {\it a)} The deformation is a continuous process, and does not
make the lattice unstable.\\
  {\it b)} The topography of the energy $E(z)$ is the simplest
required by modular symmetry: no critical points other than the
$\tl$ and $\cb$ points (see Fig.~\ref{f7} and
Sec.~\ref{sec:fareypartition}).\\
   Condition {\it a)} is stated in the theorem, and the role and
meaning of condition {\it b)} will become evident in the proof
(see Sec.~\ref{sec:lemmas}). The conditions {\it a)} and {\it b)}
define implicitly a large family of interactions, apparently
including all repulsive potentials. At the moment, however, an
explicit characterization of this family is lacking. (Maybe it
is worthwhile to mention that we failed to find exceptions among
repulsive interactions.)

The proof of the theorem relies on the modular symmetry of the
energy $E(z)$ (see Sec.~\ref{sec:modular_symmetry}),
which gives the topography of $E(z)$ in the entire $z$ plane
from that within a single Farey triangle. Also, we will use the
relation between Farey triangles and parastichy domains, already
mentioned in Sec.~\ref{spaceoflattices}. We will discuss it
again and summarize in Sec.~\ref{sec:fareypartition}. The plan of
the proof includes the following two steps. First, by making use
of the modular symmetry, we replace the {\it global} statement
about the trajectories behavior in the entire $z$ plane by an
equivalent {\it local} statement about the behavior in a single
Farey triangle (see {\it Lemmas 1, 2, 3} in
Sec.~\ref{sec:lemmas}). Then, by using modular symmetries, we
treat a minima trajectory within a single Farey triangle, and
find the relation with the parastichy domains. For that, we map
the lines $y={\rm const}$ in an arbitrary Farey triangle onto a
certain family of curves in a ``reference'' Farey triangle, and
then (locally) minimize the energy on these curves. At this
step, we establish a Fibonacci type relation between the
parastichy domains traced by a trajectory.

Loosely speaking, the Farey triangles play an ``organizing
role'' in the $z$ plane. They represent something like standard
blocks, or units for the trajectories. In some sense, the
behavior of trajectories inside all triangles is similar, and the
topology of the trajectories pattern can be constructed by
replicating one triangle with the modular symmetries.

\subsection{Modular symmetry and Farey triangles}
\label{sec:fareypartition}

Here we review basic facts about Farey
triangles on the hyperbolic plane, and study the relation of the Farey
triangles and parastichy domains. Necessarily,
our discussion be brief. (We refer
interested reader to very good texts by Apostol~\cite{Apostol}
and Iversen~\cite{hyperbolicgeometry}.)

The geometric role of modular transformations is that they
preserve the hyperbolic metric $dl^2 =|{\rm Im}\,z|^{-2}d\bar z
dz$. The ${\rm Im}\,z\,\ge\,0$ halfplane supplied with this metric
is called the {\it hyperbolic plane}. The transformations that
preserve the hyperbolic metric (they are called {\it
isometries}) form a group known as $PSL_2(R)$ or $SL_2(R)\otimes
Z_2$, that plays a role in the hyperbolic geometry similar to
that of space translations and rotations in the Euclidean
geometry. All hyperbolic isometries have the form
(\ref{eqn:modularsymmetry}) with arbitrary real $a$, $b$, $c$,
and $d$, such that ${\rm det}{\bf A}=\pm1$. The modular
transformations (\ref{eqn:modularsymmetry}) with integer $a$,
$b$, $c$, $d$, and ${\rm det}{\bf A}=\pm1$, form an infinite
discrete subgroup of the group of all isometries, as, for instance, 
symmetries of a crystal in the Euclidean space do. The
group theory notation for the modular group is $PSL_2(Z)$ or
$SL_2(Z)\otimes Z_2$.

The geodesics~\cite{hyperbolicgeometry} of the hyperbolic metric
are semicircles with the diameters on the real axis:
   \be
\label{geodesic}
y=\sqrt{\left(x-x_0\right)\left(x_1-x\right)}\ ,\ \ x_0<x<x_1\ .
   \ee
Obviously, the isometries (\ref{eqn:modularsymmetry})
map any geodesic to a geodesic. There
are two different classes of hyperbolic isometries:\\
   {\it a)} The mappings with ${\rm det}{\bf A}=+1$ that have
just two fixed points on the real axis;\\
   {\it b)} The mappings with ${\rm det}{\bf A}=-1$ that have a
whole geodesic of fixed points.\\
  Given a geodesic (\ref{geodesic}), the mapping that leaves it
fixed is given implicitly by
  \be\label{reflection}
{z'-x_0\over z'-x_1}=- {\bar z-x_0\over\bar z-x_1}\ .
  \ee
Such a transformation is analogous to a Euclidean reflection, with
the geodesic (\ref{geodesic}) corresponding to a mirror. A
composition of an even number of transformations of the type
{\it b)} is a transformation of the type {\it a)}.

The property of Farey triangles that makes them useful is that
the modular transformations (\ref{eqn:modularsymmetry}) map
any Farey triangle either onto another Farey triangle, or
onto itself. To see why is that, one first notes that by a
transformation (\ref{eqn:modularsymmetry}) with unrestricted
$a$, $b$, $c$, and $d$, we can map any three points on the real
axis onto any other three points (because it is a
fractional-linear function). Then one checks that, since
vertices of Farey triangles are rationals, the numbers $a$, $b$,
$c$, and $d$ can be chosen to be  integer, with ${\rm det}{\bf
A}=\pm1$. This follows from the explicit form of the
transformation (\ref{eqn:modularsymmetry}) that maps an
(arbitrary) Farey triangle $[p/m, q/n, (p+q)/(m+n)]$ onto the
triangle $[0/1, 1/1, 1/2]$:
    \be
{\bf A}=\left(
  \begin{array}{cc}
    m & -p \\ m-n & -p+q
  \end{array} \right)\ .
\label{eqn:examplemapping1}
   \ee

Applied to the lattice energy $E(z)$, modular symmetries relate
the values of $E$ at different points of the $z$ plane. Given
$E(z)$ inside one Farey triangle, the modular transformations
extend it throughout the whole plane. However, let us emphasize
that the function $E(z)$ within one Farey triangle has certain
symmetry properties, and hence it is not completely arbitrary.
The reason is that, as it was mentioned above, the Farey
triangles are larger than the actual fundamental domains of the
modular group. For each triangle there are six modular
transformations (including the indentity) that map it onto
itself. The energy $E(z)$ is invariant under these
transformations, and hence there is a usual relation between the
symmetries of a triangle and the behavior of the function $E(z)$
inside it:\\
  {\it a)}  There  are extremal points of $E(z)$ at all symmetry
points of the triangle transformations;\\
  {\it b)} The contours of $E(z)$ are normal  to  the  invariant
lines of the transformations.\\
  Since the topography of the energy $E(z)$ within a Farey
triangle will be crucial for our discussion, let us describe
here the triangle symmetries in some detail.

There is a total of six  modular  transformations  which  map  a
Farey triangle onto itself. From the group theory point of view,
the  symmetry  group of a Farey triangle is identical to that of
an equilateral triangle in a Euclidean plane that includes three
reflections, two rotations, and the identity.  Each  permutation
of the triangle vertices defines a modular transformation (since
it  is  a  fractional-linear  transformation).  For example, the
transformation  of  the  triangle  $[p/m,q/n,(p+q)/(m+n)]$  that
interchanges $p/m$ with $q/n$, and preserves $(p+q)/(m+n)$, is a
reflection  (\ref{reflection})  with  respect  to  the  geodesic
$[(p+q)/(m+n),(p-q)/(m-n)]$.   The   two    other    reflections
correspond    to   the   geodesics   $[p/m,(p+2q)/(m+2n)]$   and
$[q/n,(2p+q)/(2m+n)]$.  Like  in  the   Euclidean plane, where a
composition  of  two  reflections is  a rotation, for the Farey
triangle the composition of two reflections is a  transformation
giving  rise  to  a  cyclic  permutation  of  the  vertices. The
intersection of the three geodesics
    \ber
    \label{eqn:triangularxy}
z & = & {pm+qn+pn/2+mq/2 \over m^2+mn+n^2}\\
  & & +\ i\ {\sqrt{3} \over 2}{1 \over m^2+mn+n^2} . \nonumber
    \eer
is  a  symmetry  point  of  the triangle invariant under all six
symmetries of the triangle.

The symmetry lines of the Farey triangle are the three geodesics
listed  above,  and  also  the  three  sides  of  the  triangle,
$[p/m,q/n]$, $[p/m,(p+q)/(m+n)]$, and $[q/n,(p+q)/(m+n)]$, since the
reflections (\ref{reflection}) about them are modular
transformations that map the triangle $[p/m,q/n,(p+q)/(m+n)]$ to
the three adjacent triangles. There are three symmetry points on
the sides of the triangle given by the intersections of the
symmetry lines. For example, the side $[p/m,q/n]$ intersects
with $[(p+q)/(m+n),(p-q)/(m-n)]$ at
    \be\label{squarepoint}
  z={pm+qn \over m^2+n^2} +
      i{1 \over m^2+n^2}  .
    \ee

The symmetry lines and the points (\ref{eqn:triangularxy}) and
(\ref{squarepoint}) have a simple meaning in terms of the
geometry of the lattice (\ref{lattice}). The Farey triangle
sides correspond to the lattices with a rectangular unit cell.
The geodesics $[(p+q)/(m+n),(p-q)/(m-n)]$,
$[p/m,(p+2q)/(m+2n)]$, and $[q/n,(2p+q)/(2m+n)]$ correspond to
the lattices with a rhombic unit cell. One verifies this
immediately by checking that the conditions ${\bf r}_{pm}\cdot
{\bf r}_{qn}=0$, and $|{\bf r}_{pm}|=|{\bf r}_{qn}|$ define the
geodesics $[p/m,q/n]$, and $[(p-q)/(m-n),(p+q)/(m+n)]$,
respectively. Consequently, the points of the form
(\ref{eqn:triangularxy}) correspond to perfect triangular
lattices, and the points (\ref{squarepoint}) correspond to
square lattices. We will call them $\tl$ points and
$\cb$ points, respectively. Each Farey triangle contains
one $\tl$ point in its interior and three $\cb$ points,
one per each side.

The parastichy domains, as we found in
Sec.~\ref{spaceoflattices}, are defined by the lines
corresponding to rhombic lattices, i.e., by the symmetry lines.
The corners of the parastichy domains are the $\tl$ points. The
boundaries are the portions of the symmetry lines extending from
the $\tl$ point to the Farey triangle vertices. At each $\tl$
point three parastichy domains are adjacent. The parastichy
pairs of the three parastichy domains overlapping with a Farey
triangle $[p/m, q/n, (p+q)/(m+n)]$ are $(m,n)$, $(m,m+n)$, and
$(n,m+n)$.

The symmetry points (\ref{eqn:triangularxy}) and
(\ref{squarepoint}) are extremal points of the energy $E(z)$. At
the $\tl$ points the energy must have local minimum, since a
perfect triangular lattice is a lowest energy configuration for
an isotropic repulsive interaction. The minima at the centers of
the valleys inside the Farey triangles in Fig.~\ref{f7} are at
the $\tl$ points. At the corners of the Farey triangles the
energy is maximal, since at these points one of the lattice
periods vanishes, and the energy diverges. Given that the maxima
are at the corners, and the minima at the centers, the $\cb$
points have to be the saddle points. (The square lattice has
higher energy than the triangular lattice, and it is unstable
with respect to a deformation towards triangular lattice through
a continuous sequence of rhombic lattices.) Such a configuration
of extremal points, located only at the symmetry points, is a
minimal combination required by symmetry. For an arbitrary
interaction, in principle, one could have other extremal points
of the energy, not associated with the symmetry points of the
Farey triangles. However, that would make the discussion
unnecessarily complicated, and moreover, for typical simple
repulsive interactions (exponential, gaussian, power law, etc)
there are no additional extremal points. Therefore, in the
following discussion we will assume that there are no other
extremal points besides those required by symmetry.

Another useful property is that, by symmetry, the contours of
the energy $E(z)$ are normal to the Farey triangle sides and
the symmetry lines, everywhere except the $\cb$ and $\tl$ points.
This result will be crucial in the following discussion.

\subsection{Farey partition of the hyperbolic plane and the trajectories of energy minima}
\label{sec:lemmas}

To prove the main theorem, we have to see why the whole infinite
sequence of the parastichy domains traced by an energy minima
trajectory has the parastichy pairs that obey the Fibonacci
addition rule. For that we will study the relation between the
Farey triangles and the minima trajectories. The theorem
statement is {\it global}, since it states the trajectories
behavior in the entire $z$ plane. Our first step will be to
replace it by {\it local} statements which are equivalent to the
theorem, but talk only about one Farey triangle. One can say
that we reduce the mystery all over the $z$ plane to the mystery
inside one Farey triangle. The following three Lemmas together
are equivalent to the statement of the main theorem.

\noindent
{\bf Lemma 1\ } The minima trajectories enter or leave Farey
triangles only through the $\cb$ points.

\noindent
{\bf Lemma 2\ } The trajectories enter or leave the parastichy
domains only through the $\tl$ points. Any $\tl$ point belongs
to one of the trajectories.

\noindent
{\bf Lemma 3\ }
At a $\tl$ point, by going from one to another parastichy domain
the trajectory obeys the Fibonacci rule: after exiting the
domain $(m,n)$, $m<n$, it enters the domain $(n,m+n)$.

Given the starting point of a continuous trajectory  of  minima,
the  three Lemmas completely determine through which sequence of
Farey triangles and parastichy domains the trajectory  proceeds.
The  {\it  Lemmas  1},  {\it  2}  state  that,  no  matter  what
interaction,  there  are  $\cb$  and  $\tl$   points   on   each
trajectory.  Although  the  trajectory  may  be sensitive to the
interaction, the sequence of the $\cb$ and $\tl$  points  through
which  it  goes  is  decided  purely geometrically, and does not
depend on the interaction.

As for the $\tl$ points, since perfect triangular lattice has
lowest energy, it is understandable that the $\tl$ points always
lie on trajectories. However, the fact that (some) $\cb$ points
also lie on the trajectories comes as a surprise. We will see
that both {\it Lemma 1} and {\it 2} follow from the modular symmetry.

{\it Lemma 3} states that the behavior of trajectories near the
$\tl$ points determines the evolution of parastichy pairs. Each
$\tl$ is a corner of three parastichy domains (say, with the
pairs $(m,n)$, $(m,m+n)$, and $(n,m+n)$). The rule that the
change of the pair is always $(m,n)\rightarrow (n,m+n)$, where
$m<n$, applied iteratively along the trajectory, evidently leads
to a generalized Fibonacci sequence. According to {\it Lemma 3},
there is no trajectory branching: new trajectory in the
parastichy domain $(m,m+n)$, $m<n$, emerges at a finite distance
from the $\tl$ point.

Note that {\it Lemma 3} is not applicable to the triangle
$[0/1,1/1,1/2]$, ($m=n=1$), where branching occurs at the
bifurcation point.

\noindent{\bf Proof:}\hskip5mm
To  prove  the  lemmas  for  a   given   Farey   triangle   (say
$[p/m,q/n,(p+q)/(m+n)]$, $m<n$) we use modular symmetry
(\ref{eqn:examplemapping1}),
and map this triangle onto $[0/1,1/1,1/2]$.
To  study  how the
energy minima at constant $y$ transform under this  mapping,  we
have  to  find  out how  the  constraint $y={\rm const}$ transforms.
Under the transformation (\ref{eqn:examplemapping1}), the family
of lines $y=\lambda$ is mapped to a family of circles tangent to
the real axis,
    \be
|z-(x_0+i y_0)|^2=y_0^2,
\label{eqn:ycont1}
   \ee
where  $x_0=m/(m-n)$,  and $y_0=1/(2(m-n)^2\lambda)$. The images
of the minima can be found by (locally) minimizing the energy on
the circles (\ref{eqn:ycont1}).

\begin{figure}[t]
\includegraphics[width=3.5in]{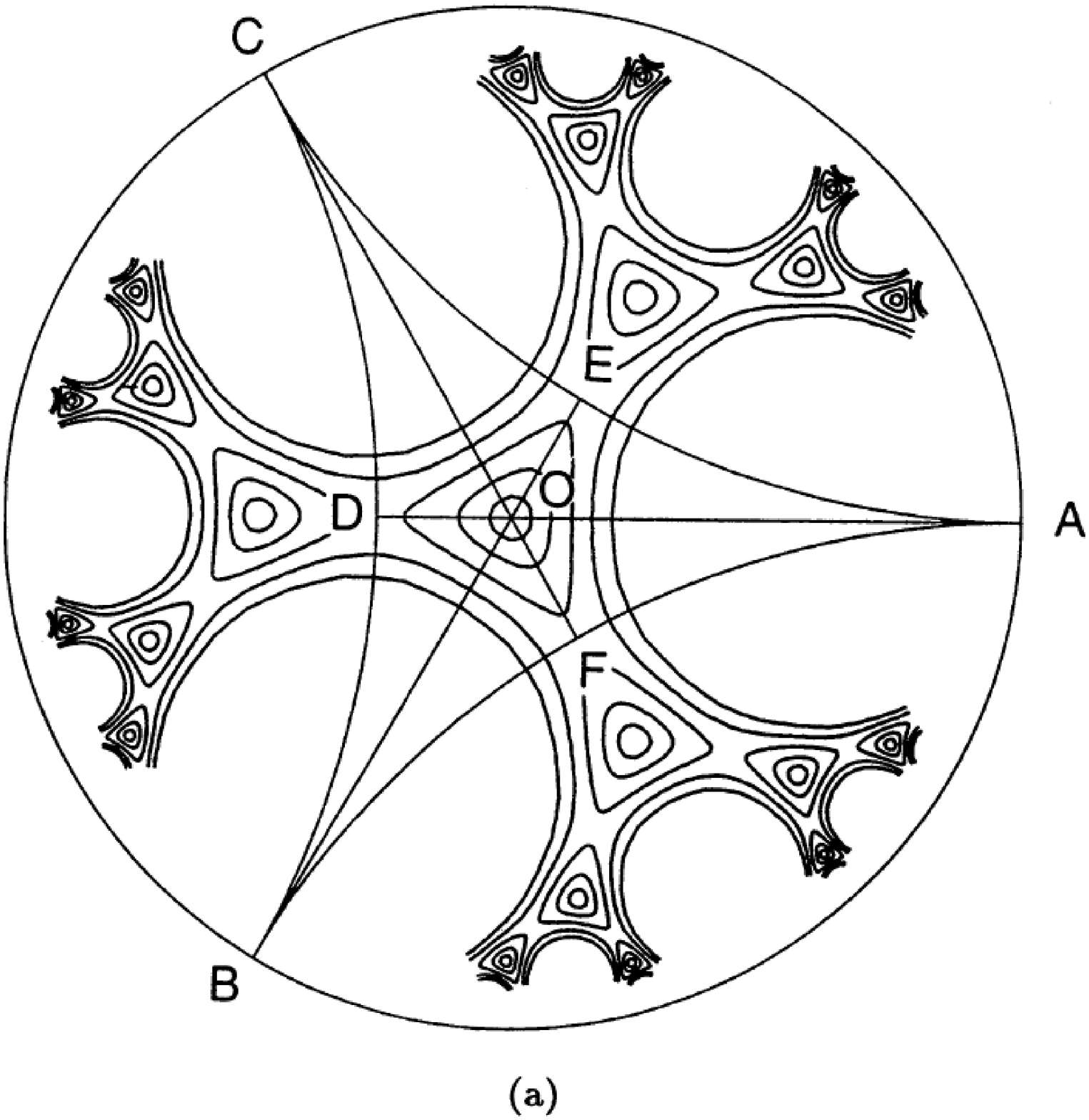}
\includegraphics[width=3.5in]{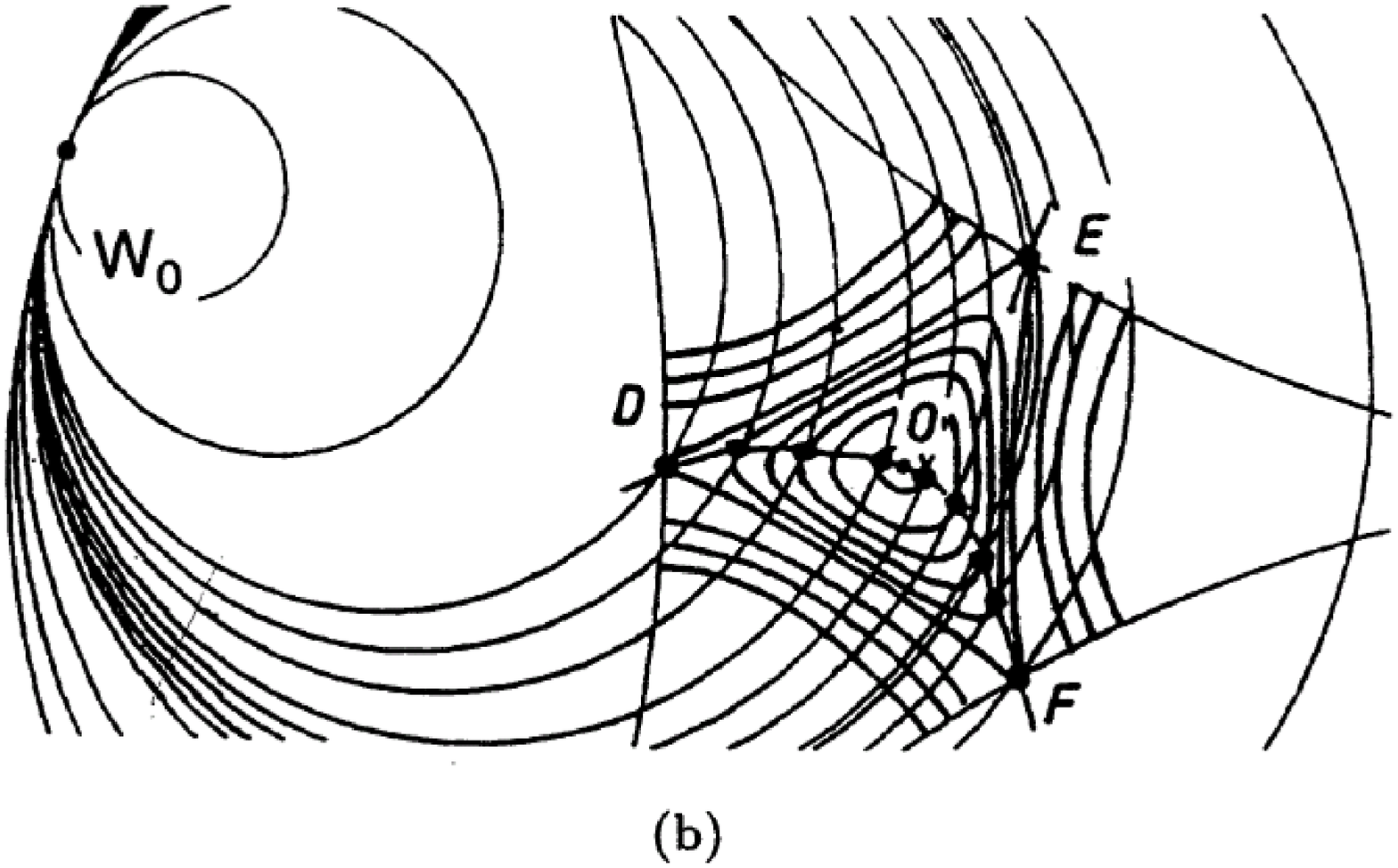}
\caption{
  {\it (a)} A contour plot (thick lines) of the  lattice  energy
for the interaction $U(r)=\exp(-5r)$, mapped to the $w$ plane by
(\protect\ref{eqn:extramapping}).  The  triangle  $[C,B,A]$ is a
mapping of the Farey triangle $[0/1,1/1,1/2]$.
  {\it (b)} A  trajectory  of  energy  minimum  enters  the  triangle
$[A,B,C]$  through  the  point  $D$, and passes the point $O$ to
exit through the point $F$. A new trajectory is created near the
point $E$, and exits through the point $E$.}
\label{f8}
\end{figure}

This task is  greatly  facilitated  by  the  following  variable
change,
   \be
z \rightarrow w=(-i\sqrt{3}\overline{z}+\epsilon)/
     (i\sqrt{3}\overline{z}+\overline{\epsilon})  ,
\label{eqn:extramapping}
   \ee
where $\epsilon=e^{i\pi/3}=1/2+i\sqrt{3}/2$. It maps the upper  $z$  plane
onto  a  unit circle $|w|=1$. Fig.~\ref{f8}a shows the contours
of the energy taken as a function of $w$.

Under the variable change (\ref{eqn:extramapping}), the Farey
triangle $[0/1,1/1,1/2]$ maps onto the triangle
$[\epsilon^2,\overline{\epsilon}^2,1]=[C,B,A]$. The $\tl$ point
maps to the point $O$ and the three $\cb$ points map to the
points $D$, $E$, and $F$. The parastichy boundaries map onto the
lines $OA$, $OB$, and $OC$. The curved triangles $AOB$, $BOC$,
$COA$ belong to parastichy domains $(n,m+n)$, $(m,n)$, and
$(m,m+n)$, respectively.

The family of circles in Eq.~\ref{eqn:ycont1}
maps to a new family of circles
\ber
 & |w-w_0(1-r)|=r, & \nonumber \\
 & w_0=(m-(m+n)\epsilon)/(m-(m+n)\overline{\epsilon}),
    \label{eqn:ycont2} & \\
 & r=\sqrt{3}/\left(\sqrt{3}+2\lambda
    \left(m^2+mn+n^2\right)\right).
    \nonumber &
\eer
which are tangent to  the  unit  circle  at  $w_0$.  We  compute
  \be
{\rm arg}(w_0)=2\,  \arctan\left(n+m)/(n-m)\right),
  \ee
and
find $2\pi/3<\mbox{arg}(w_0)\le \pi$, since $0\le m\le n$.

The  behavior  of  the  energy  contours  within  the   triangle
$[A,B,C]$ follows from the results of Sec.~\ref{sec:modularsymmetry}.
The  contours  are  smooth curves everywhere except the extremal
points: $A$, $B$, $C$, $D$, $E$, $F$, and $O$.  Also,  according
to the last statement of Sec.~\ref{sec:modularsymmetry}, the contours
are  everywhere  perpendicular to the arcs $AB$, $BC$, $CA$, and to the
segments $AD$, $BE$, $CF$.

Now we are ready to study the trajectories.  Geometrically,  the
trajectories  points are those where the contours are {\it tangent} to
the circles (\ref{eqn:ycont2}).
It is simple to see that the circles  (\ref{eqn:ycont2})  cannot
be perpendicular to $AB$, $BC$, $CA$, $AD$, $BE$, $CF$, provided
that  $\mbox{arg}(w_0)\ne \pi$. But
$\mbox{arg}(w_0)=\pi$ can occur only for
the special
case $m=1$, $n=1$, excluded by the condition $m<n$.
Thus, we proved
{\it Lemma 1} and {\it Lemma 2}.

Modular  symmetry implies that in the vicinity of
the $\tl$ point $O$ the contours are  circles.  From  that,  it
follows  that  at  the  point $O$ the trajectory is tangent to the
straight line joining $O$ and $w_0$  (Fig.~\ref{f8}(b)),  and  also
that   the   trajectory   branching   does   not  happen.  Since
$2\pi/3<\mbox{arg}(w_0)<\pi$, the trajectory in $BOC$ passes $O$
to enter $AOB$. Recalling that parastichy  pairs  in  $BOC$  and
$AOB$  are  $(m,n)$ and $(n,m+n)$, respectively, one gets
the Fibonacci rule, thus proving {\it Lemma 3}.
\hfill {\bf QED}

Finally, let us discuss the case $m=1$, $n=1$, in which {\it
Lemma 3} does not hold. However, if the method of the proof is
applied to this case, it yields $\mbox{arg}(w_0)=\pi$, that is a
family of circles (\ref{eqn:ycont2}) perpendicular to $AD$.
Then, obviously {\it Lemma 1} is still valid, since the circles
cannot be perpendicular to $AB$, $BC$, $CA$. The trajectory goes
along the segment $AD$ up to some point within the segment $AO$,
where the bifurcation occurs. At this point, the trajectory
gives rise to a pair of symmetric principal trajectories, which
reflects the $x\rightarrow 1-x$ symmetry. Thereafter, the
principal trajectories are described by the lemmas, and hence
correspond to Fibonacci structures.

\section{Robustness of the problem}
\label{sec:robustness}
\subsection{Deformation {\it versus} anisotropic growth}
\label{subsec:cylindricalphyllotaxis}
   So far, we  studied  cylindrical  lattices  under  the  fixed
density  condition (or, equivalently, under fixed pressure), and
found a relation between phyllotaxis and the deformation  caused
by  a  uniaxial stress applied along the cylinder axis. However,
in order to understand the ubiquity of phyllotaxis, as discussed
in Sec.~\ref{subsec:growthandstress},  one  has  to  apply  these
results  to  the growth in different geometries: cylinder, disk,
or cone. In this section we will  discuss  implications  of  the
theorem   proved  in  Sec.~\ref{sec:modular_symmetry}  for  these
growth problems. We will see that, because of the robustness  of
the  problem,  phyllotaxis  can be understood in each case by an
appropriate modification of the deformation picture.

Let us start with the simplest case of a cylindrical structure
that grows anisotropically: faster in the cylinder crossection,
and slower along the axis. Intuitively, up to a dimension
rescaling by a factor $\sqrt{\rho}$, where $\rho$ is the density,
we return to the deformation problem at fixed density. However,
since the rescaling changes the interaction $U(r)$, and this
might affect the growth, let us look at the problem closer.

The   anisotropic   growth   of    a    cylindrical    structure
(\ref{spiral_lattice}) can be accounted for by assuming that the
density   and   the  interaction  are  some  functions  of  $y$:
$\rho(y)$, $U_{y}(r)$, since  decreasing  $y$  in  this  problem
plays  a  role of the time. By density rescaling, the energy $E$
can be brought to the form in which  $A=\rho^{-1}=1$,
and the interaction parametrically depends on $y$:
  \ber
    E(x,y;A,U(r)) & = & E(x,y;1,U_{eff}(r)), \nonumber \\
    U_{eff}(r) & = & U_{y}(\sqrt{A}r).
\label{eqn:scalerelation}
  \eer
This  formula  replaces  the  density  change  by  an  effective
interaction evolution. To find out the effect of the  latter  on
the energy minima trajectories, one has to go back to {\it Lemma
1,\  2,}  and  {\it  3}.  Obviously,  as  long  as the effective
interaction   in    Eq.~\ref{eqn:scalerelation}    satisfy    the
requirements  listed at the end of Sec.~\ref{sec:discussion}, the
lemmas hold:  the  minima  trajectories  enter  and  exit  Farey
triangles  through  the  $\cb$  points;  they enter and exit the
parastichy domains through the $\tl$ points; and  at  the  $\tl$
points  the Fibonacci rule is obeyed. The reason is that each of
the lemmas is a ``local'' statement with respect to $y$,  as  it
can  be  verified by considering the situation at one particular
$y$.

From  that,  we  conclude  that  the  deformation and the growth
problems are equivalent. By a similar argument, one  can  extend
the  result  to  the case when the interaction varies during the
growth in some arbitrary way,  consistent  with  the  conditions
required by the theorem.

Finally, there are interactions for which  the  deformation  and
the   growth   problem   are   equivalent   exactly,   not  just
topologically, or in the sense of parastichy  pairs.  Any  power
law  potential  $U(r)=U_0r^{-\gamma}$  has such a property, which
follows from its scale invariance.

\subsection{The disk and cone geometries}
\label{subsec:spiral}
   Plants with disk geometry exhibit spiral phyllotaxis (see
Fig.~\ref{f1}(b)), which apparently is quite similar to the
cylindrical phyllotaxis. Naively, one can say that a complex
mapping $z\rightarrow\ln z$ transforms one problem to the other,
as it maps punched $z$ plane to a cylinder, and transforms
log--spirals in the $z$ plane to helices on the cylinder.
However, there is one interesting feature that distinguishes the
spiral phyllotaxis from the cylindrical one. For example, a
sunflower is divided into circular concentric domains of ring
shape which have different parastichy numbers. From one domain
to another, the parastichy numbers change so that they grow
outwards the disk, and the transition of the numbers across the
domain boundary (called ``parastichy transition'') follows the
Fibonacci rule.

To  apply  the  energy model, we note that sufficiently far away
from the center, outside a core region,  the  structure  locally
looks  like a lattice. So, one can write a lattice energy, as we
did for the cylindrical problem, and study the evolution under a
stress. The origin of the stresses in this case, as  we  discussed
in  Sec.~\ref{subsec:growthandstress},  is  at  the  disk center,
where during the growth new structure units are being generated.
Older units are pushed by newer ones, and move outwards  as  the
structure grows.

Let us model the structure by the points of a spiral lattice
(Fig.~\ref{f1}(b)), and order the lattice points,
given by $(\rho_i, \theta_i)$ in cylindrical coordinates, 
in the order of increasing radius $\rho_i$. 
To characterize the structure locally as a periodic
lattice, in analogy with the discussion
in Sec.\ref{sec:history} (cf. Eq.(\ref{spiral_lattice}), we introduce two sets of parameters:
  \be
{\sf r}_i = \rho_{i+1}-\rho_i \, , \quad
a_i =  2\pi \rho_i \, , \quad
{\sf d}_i = \theta_{i+1}-\theta_{i} \, ,
   \label{eqn:spiralparameters1}
\ee
and
\be
x_i = -{\sf d}_i / 2 \pi , \quad
y_i = {\sf r}_i / a_i , \quad
A_i = {\sf r}_i a_i .
  \label{eqn:spiralparameters2}
  \ee
We  assume  that  outside a core region, $i>i_{\rm min}$, the parameters
are slowly varying as function of $i$, so  that  the  notion  of
local   periodic   lattice can be used. The parameters
(\ref{eqn:spiralparameters1}) and  (\ref{eqn:spiralparameters2})
have   the   same   geometrical  meaning  as  the  corresponding
parameters   (\ref{spiral_lattice}),   (\ref{lattice})   of    a
cylindrical lattice.

The density is approximately constant  throughout  the  disk,
outside   the   core   region.   In   terms  of  the  parameters
(\ref{eqn:spiralparameters1}) and (\ref{eqn:spiralparameters2}),
$A_i=A={\rm const}$ for $i>i_{\rm min}$. From  that,  outside  the  core
region,   $y_i$   decreases   with  the  radius:  $y_i=A/(2\pi
\rho_i)^2$. Then, since $A_i$ and $y_i$ are completely fixed  by
the  growth  process, $x_i$ is the only free parameter left, and
we again have a problem of optimizing $x_i$ so that the  lattice
energy attains a local minimum.

For  a  short  range  interaction  of  radius $\rho_0$, the size of the core
region $\rho_{\rm c}$ti is comparable to $\rho_0$.  Inside  this
region one cannot identify periodic lattice, even approximately,
and  the growth has to be described in a different way. However,
the study of the core region  lies  beyond  the  scope  of  this
paper,  and we refer the reader to other chapters of this volume
as well as   
Refs.~\cite{Mitchison,Meinhardt,CouderDouady}.  
Below  we  
assume  that,  due to the short
range of interaction, the core has essentially no effect on  the
lattice  structure far away, where parastichy transitions occur.
The role  of  the  core  is  merely  in  setting  right  initial
conditions  for  the  deformation process, which can be compared
with  the  selection  of  initial  lattice  in  the  cylindrical
problem. Similarly to that problem, as we will see, once initial
structure   is   given,   the final state of  its  deformation  is
determined anambiguously.

Now, let us discuss the growth  process.  Total  energy  of  the
structure is given by
  \be
E_{\rm tot} = \sum_{i>i_c} E_i \, ,\quad
E_i = {1 \over 2}\sum_{j \neq i}
  U(|{\bf r}_j-{\bf r}_i|),
  \ee
where  $E_i$  is  the  energy  of the point $i$ interacting with
other points. For  a  short  range  interaction,  $E_i$  can  be
replaced  by  the  energy $E(x_i,y_i;A_i)$ of a periodic lattice
whose structure is locally  identical  to  that  of  the  spiral
lattice,
  \be
E_{\rm tot}=\sum_{i>i_c} E(x_i,y_i;A_i) \, .
  \ee
For  slowly  varying  parameters,  the summation over $i$ can be
replaced by an integration over $\rho$,
   \be
E_{\rm tot}=\int_{\rho_c} \, d\rho \, 2\pi\rho \,
  A^{-1}(\rho) \, E(x(\rho),y(\rho);A(\rho)) \, ,
   \ee
where $2\pi\rho  \,  A^{-1}(\rho)$  is  the  radial  density.  We note  that $E_{\rm tot}$ depends only on
$x(\rho)$, since $y(\rho)$ and $A(\rho)$ are fixed.

To  find  a  stable  structure  of  a spiral lattice, one has to
minimize $E_{\rm tot}$ with respect to $x(\rho)$, with a  constraint
that  $x(\rho)$  is  continuous.  Evidently,  this  minimization
problem  leads  to  the  same  minima  trajectories  as  in  the
cylindrical  problem,  since  $\rho$ and $y$ are in a one-to-one
relation. This equivalence fixes the solution of the disk problem 
$x(\rho)$, 
making the sequence of parastichy pairs in the  two
problems identical.

One  can say that the time of the cylindrical problem became the
radial dimension of the disk problem.  The  reason  is  that  the
degree  of  compression  $y$  in  the  disk problem decreases as
$\rho^{-2}$  at  $\rho\ge\rho_0$.  Hence,  the   change   of   a
parastichy pair is represented in the disk problem by a circular
boundary at which a parastichy transition occurs.

Finally, we  comment  on  the  universality  of  the parastichy
transitions.  Evidently,  the  particular  form of $A(\rho)$ and
$y(\rho)$  assumed  above  can be replaced basically by any other reasonable dependence, by an argument similar to that of
Sec.~\ref{subsec:cylindricalphyllotaxis}.     The      parastichy
transitions  are therefore a  universal  property  of  spiral lattices,
independent  of  the details of the density  profile.  
In this way our discussion  of  parastichy
transitions can be extended to other geometries, e.g. the cone geometry intermediate between those of a  disk
and cylinder, supplying a general argument for the stability of Fibonacci numbers.

\section{Summary}
\label{subsec:summary}
   The mechanical theory explains phyllotaxis by stipulating 
that the growth of a plant at 
a critical stage
(perhaps embryonal) is anisotropic. 
It argues that stress buildup and relaxation in a deformable lattice
induced by its growth is such that it leads to
Fibonacci  phyllotactic patterns, exclusively     and
deterministically.  In  the  mechanical  theory reviewed above this  result is
derived rigorously 
by analyzing the mechanics of deformable cylindrical lattices. 
This theory explains the predominance of  Fibonacci  numbers  in
phyllotaxis,  as well as why the 
most frequent exceptions in phyllotaxis are described by Lucas numbers.

The robustness of 
phyllotactic growth in other geometries, such as a disk or a cone, can be understood by combining the stability if Fibonacci numbers in cylindrical 
lattices with the general properties of deformable lattices growing under stress. Indeed, stress buildup and relaxation in non-cylindrical lattices, if viewed locally, follow the same rules as in the cylindrical case. 
As a result, the outcome of the
anisotropic deformation is insensitive to  the  specifics 
of the interaction energy of the phyllotactic pattern or its geometry, be it a
cylinder, a disk, or a cone. This provides further insight into
the ubiquity of Fibonacci numbers in natural growth.



\end{document}